\documentclass{aa}  
\usepackage{graphicx}
\usepackage{txfonts}
\usepackage[dvipsnames]{xcolor}
\usepackage[normalem]{ulem}
\usepackage{array, multirow, boldline}
\usepackage{multirow}
\usepackage{caption}
\usepackage{threeparttable} 
\usepackage{booktabs} 

\usepackage[colorlinks,linkcolor=MidnightBlue,citecolor=MidnightBlue,linktocpage=true,breaklinks, 
plainpages=false,urlcolor=MidnightBlue]{hyperref}
\def\thethree{{\sc The Three Hundred Project}}

\usepackage[]{subfig}
\begin{document}

   \title{Deriving accurate galaxy cluster masses using X-ray thermodynamic profiles and graph neural networks}


   \author{Asif Iqbal
          \inst{1,2}
          \and 
          Subhabrata Majumdar\inst{2}
          \and
          Elena Rasia\inst{3,4,5}
          \and 
          Gabriel W. Pratt \inst{6}  
          \and 
          Daniel de Andres\inst{7}
          \and
          Jean-Baptiste Melin\inst{8}
          \and
          Weiguang Cui\inst{9}
          }
   \institute{Univ. Lille, Univ. Artois, Univ. Littoral C\^ote d'Opale, ULR 7369 -- URePSSS -- Unit\'e de Recherche Pluridisciplinaire Sport Sant\'e Soci\'et\'e,  F-59000 Lille, France
              \email{asif-iqbal.ahangar@univ-lille.fr}
         \and
        Tata Institute of Fundamental Research, 1 Homi Bhabha Road, Colaba, Mumbai 400005, India
             \email{subha@tifr.res.in}
           \and 
           INAF - Osservatorio Astronomico di Trieste, via Tiepolo 11, I-34131 Trieste, Italy
           \and IFPU, Via Beirut, 2, 3I-4151 Trieste, Italy
           \and
           Department of Physics; University of Michigan, Ann Arbor, MI 48109, USA
        \and 
        Universit\'e Paris-Saclay, Universit\'e Paris Cit\'e,  CEA, CNRS, AIM, 91191, Gif-sur-Yvette, France 
        \and
        Nonlinear Dynamics, Chaos and Complex Systems Group, 
        Departamento de Biología y Geología, Física y Química Inorgánica, 
        Universidad Rey Juan Carlos, 
        Tulipán s/n, 28933 Móstoles, Madrid, Spain
        \and
         Universit\'e Paris-Saclay, CEA, D\'epartement de Physique des Particules, 91191, Gif-sur-Yvette, France
         \and 
         Departamento de F\'isica Te\'{o}rica, M\'{o}dulo 15 Universidad Aut\'{o}noma de Madrid, 28049 Madrid, Spain
             }

   
   \titlerunning{Galaxy clusters masses from X-rays using deep learning}

 
  \abstract
{Precise determination of galaxy cluster masses is crucial for establishing reliable mass-observable scaling relations in cluster cosmology. We employ graph neural networks (GNNs) to estimate galaxy cluster masses from radially sampled profiles of the intra-cluster medium (ICM) inferred from X-ray observations. GNNs naturally handle inputs of variable length and resolution by representing each ICM profile as a graph, enabling accurate and flexible modeling across diverse observational conditions. We trained and tested GNN model using state-of-the-art hydrodynamical simulations of galaxy clusters from  \thethree. The mass estimates using our method exhibit no systematic bias compared to the true cluster masses in the simulations. Additionally, we achieve a scatter in recovered mass versus true mass of about 6\%, which is a factor of six smaller than obtained from a standard hydrostatic equilibrium approach. 
Our algorithm is robust to both data quality and cluster morphology and it is capable of incorporating model uncertainties alongside observational uncertainties. Finally, we apply our technique to XMM-Newton observed galaxy cluster samples and compare the GNN derived mass estimates with those obtained with $Y_{\rm SZ}$-M$_{500}$ scaling relations. Our results provide strong evidence, at  5$\sigma$ level, for a mass-dependent bias in SZ derived masses, with higher mass clusters exhibiting a greater degree of deviation. Furthermore, we find the median bias to be $(1-b)=0.85_{-0.14}^{+0.34}$, albeit with significant dispersion due to its mass dependence. This work takes a significant step towards establishing unbiased observable mass scaling relations by integrating X-ray,  SZ and optical datasets using deep learning techniques, thereby enhancing the role of galaxy clusters in precision cosmology.}

   \maketitle
%

\section{Introduction}
Galaxy clusters serve as crucial probes for understanding the overall dynamics and evolution of the cosmic web \citep{1985Natur.317..595F}.  
A significant fraction of their mass resides in the enigmatic form of dark matter, an invisible and non-luminous quantity that constitutes approximately 80\%-90\% of the total matter content in these objects. The second most important component by mass is the intra-cluster medium (ICM), a hot ionised gas that emits in the X-ray via bremsstrahlung emission. The interplay between the visible and invisible components plays a pivotal role in shaping the cosmic web and in influencing the evolution of galaxies \citep{2012ARA&A..50..353K,2019SSRv..215...25P}.

Precise galaxy cluster mass estimates allow us to calibrate scaling relations, such as the mass-luminosity and mass-temperature relations derived using X-rays \citep{{lov22,kay22}}, which are essential for leveraging the cluster population and its evolution to infer the properties of the underlying cosmological model \citep{2009ApJ...692.1033V, 2010MNRAS.406.1759M,2019SSRv..215...25P}. These measurements contribute to improving the accuracy of cosmological parameters, such as the cosmological density of dark matter and dark energy, which in turn refine our understanding of the expansion history of the Universe and the associated formation of structure  \citep{2004ApJ...613...41M,2011ARA&A..49..409A,2024A&A...682A.148F}.

Galaxy cluster masses can be estimated in several ways: using galaxy orbits; via X-ray observations or the Sunyaev-Zeldovich effect, assuming hydrostatic equilibrium; or through the gravitational lensing of background galaxies. Each of these methods has its own set of challenges and limitations \citep[see e.g.][for a review]{2019SSRv..215...25P}. The hydrostatic equilibrium approach, for instance, assumes that the ICM is in perfect hydrostatic balance, which may not always be the case due to complex cluster dynamics and non-thermal pressure support; moreover, the associated X-ray measurements can be biased due to the presence of substructures and multi-temperature gas \citep{2012NJPh...14e5018R,2014ApJ...792...25N}. Gravitational lensing provides a more direct measure of the (projected) total mass, including dark matter, but requires high-quality data and sophisticated modeling to accurately interpret the lensing signals \citep{2013arXiv1312.5981H, 2014ApJ...795..163U}, in particular considering the triaxiality of the halo shape \citep[e.g.][]{EuclidCollaboration2024}. 

Different approaches have been used to estimate the potential mass bias due to the hydrostatic approximation. For example, \cite{2019A&A...621A..40E}, using a radial functional form for the  non-thermal pressure along with an universal baryonic fraction approximation, found a mass bias\footnote{Following the usual convention, we define mass bias as \(\textrm{M}^{\mathrm{HSE}} / \textrm{M}^{\mathrm{True}} = (1 - b)\), where \(\textrm{M}^{\mathrm{HSE}}\) and \(\textrm{M}^{\mathrm{True}}\) are the hydrostatic and true masses of galaxy clusters, respectively.}  of \((1 - b) = 0.87\pm0.05\) for the Planck SZ masses with a slight mass dependence. \citet{2024A&A...690A.238A} reported a bias of $(1-b) = 0.76 \pm 0.04$ and $0.89 \pm0.04$, obtained by comparing weak lensing data with mass estimates from XMM-Newton and Chandra X-ray datasets, respectively, illustrating the dependence of the measured hydrostatic bias on X-ray satellite  calibration. Similarly, \cite{2024A&A...682A.147M} comparing mass profiles extracted from the X-ray and lensing data found a hydrostatic-to-lensing mass bias of \((1 - b) = 0.73 \pm0.07\). Considering CMB cosmology, \cite{2018MNRAS.477.4957B}  using SZ power spectrum analysis estimated a mass bias of  \((1 - b) = 0.58 \pm 0.06\). Similarly, \cite{2019A&A...626A..27S} found hints of a redshift-dependent bias of $(1-b) = 0.62\pm0.05$.


In recent years, machine learning techniques have emerged as powerful tools for advancing the precision of galaxy cluster mass measurements. \citet{Ntampaka} introduced an algorithm based on Support Distribution Machines to reconstruct dynamical cluster masses. Similarly, \citet{2019Ho, 2020Ramanah, 2021ApJ...908..204H, 2022NatAs...6..936H} employed Convolutional Neural Networks (CNNs) to infer cluster masses using relative line-of-sight velocities and projected radial distances of galaxy pairs. \cite{2022NatAs...6.1325D} applied a CNN model, trained on {\thethree} simulations, to estimate M$_{500}$ for Planck SZ maps, while \citet{2022arXiv220712337F} adopted a hybrid approach combining autoencoders and random forest regression; both these studies achieving approximately 10\% scatter in reconstructing 3D gas mass profiles and total cluster mass. \citet{2023MNRAS.524.3289H} developed convolutional deep‐learning models to infer galaxy‐cluster masses from mock eROSITA X‑ray photon maps, achieving a mass scatter of 17.8\% with single‐band data, improved to 16.2\% when using multichannel energy bands, and further reduced to 15.9\% by incorporating dynamical information. Using multi-wavelength maps, \citet{2024MNRAS.528.1517D} applied a U-Net architecture to predict projected total mass density maps from hydrodynamical simulations. Additionally, \citet{2019ApJ...876...82N} and \citet{2024A&A...682A.132K} developed CNN-based models to estimate galaxy cluster masses directly from X-ray photon data. Such machine learning innovations hold great promise for tackling the challenges of cluster mass estimation and pushing the boundaries of cosmological research.

In this work, we propose a novel approach that leverages graph neural networks (GNNs) \citep{2016arXiv160902907K} to estimate galaxy cluster masses,  M$_{500}$\footnote{M$_{500}$ represents the total mass within the radius \( \text{R}_{500} \) of the sphere, whose average density is 500 times the critical density of the universe. Mathematically, M\(_{500} = \frac{4}{3} \pi \text{R}_{500}^3 \times 500\, \rho_{\text{crit}}(z) \), where \( \rho_{\text{crit}} \) is the critical density of the universe at the cluster redshift $z$.}, from X-ray-derived spherically symmetric radial thermal profiles of the ICM. By using GNNs, we effectively model the X-ray-derived ICM quantities data as a graph, allowing us to capture the intricate spatial relationships and hierarchical structures within the data. 
While \cite{2022NatAs...6.1325D,2024MNRAS.528.1517D} and \cite{2024A&A...682A.132K}  used direct observational images such as X-ray, SZ, and optical maps to estimate mass maps and consequently M$_{500}$ using a deep learning approach, our method adopts an indirect approach. In our case, our model relies on processed X-ray data, specifically the 3D deprojected radial profiles of density, temperature, and pressure. GNNs provide an advantage by being agnostic to the dimensionality of the input data, allowing them to seamlessly handle diverse input formats and shapes.

Our method uses state-of-the-art hydrodynamical simulations from \thethree, which provide a realistic and comprehensive dataset for training and testing our model \citep{cui2018}.  Previously, we have showed that this data set matched real observations and was used to develop a deep learning technique for the deprojection and deconvolution of observed (projected) temperature profiles \citep{2023A&A...679A..51I}. In this work, after calibrating our deep learning technique on simulations, we apply it to a sample of objects observed with XMM-Newton, highlighting its potential for real-world applications. Finally, we compare the M$_{500}$ values derived from the GNN with those determined using X-ray and SZ scaling relations, paying particular attention to insights gained from the application of our method to the biases in the mass estimates.


The remainder of this paper is organised as follows: in Sect. 2, we describe the data and simulation setup; in Sect. 3, we detail the GNN model and training procedure; in Sect. 4, we present the results of our mass estimation method and compare it with traditional approaches; in Sect. 5, we apply our GNN model to two X-ray samples of galaxy clusters and in Sect. 6, we discuss the implications of our findings and future directions for research. 
Throughout this work, we assume a flat $\Lambda$CDM model with $H_{0}=67.7$ km s$^{-1}$ Mpc$^{-1}$, $\Omega_{m}=0.3$, and $\Omega_{\Lambda}=0.7$. Additionally, $E(z)$ represents the ratio of the Hubble constant at redshift $z$ to its present value, $H_{0}$.

\section{Simulations}
\label{sec:simulations}
In this study, we train the neural network using the gas and dark matter radial profiles of galaxy clusters from the \thethree, dataset \citep{cui2018,2020A&A...634A.113A}. In Section~\ref{sec:sub0}, we provide a brief overview of the \thethree\, and in Section~\ref{sec:sub1} we discuss the dynamical state of the galaxy clusters. The masses derived from our GNN model will be compared to the hydrostatic masses, which are computed in Section~\ref{sec:sub2}.
\subsection{Simulated cluster sample}
\label{sec:sub0}
The cluster samples used in this study were simulated using the GADGET-X code \citep{2016MNRAS.455.2110B}, and are based on 324 Lagrangian regions centered on the most massive galaxy clusters at $z=0$ selected from the MultiDark dark-matter-only MDPL2 simulation \citep{Klypin2016}. Both the parent box and resimulations assumed the cosmological parameters from the Planck mission \citep{2016A&A...594A..13P}. The MDPL2 simulation is a periodic cube with a comoving size of 1.48 Gpc containing $3840^3$ dark matter particles. The selected regions were resimulated with the inclusion of baryons and consider the following processes for the baryonic physics: metallicity-dependent radiative cooling, the effect of a uniform time-dependent UV background, a sub-resolution model for star formation from a multi-phase interstellar medium, kinetic feedback driven by supernovae, metal production from SN-II, SN-Ia, and asymptotic-giant-branch stars, and AGN feedback \citep{2015ApJ...813L..17R}.

We analysed three samples of galaxy clusters identified at $z = 0.07$, $z = 0.33$, and $z = 0.95$. The clusters were selected within a mass range of $ 1 \times 10^{14}\, \textrm{M}_\odot \lesssim \textrm{M}_{500} \lesssim 2 \times 10^{15}\, \textrm{M}_\odot$. We included both the central clusters and any other object, in the resimulated regions, whose mass falls in this range. This leads to a total sample size of 1655 clusters, distributed as follows: 463 clusters at $z = 0.07$, 639 clusters at $z = 0.33$, and 553 clusters at $z = 0.95$.

We considered spherically symmetric 3D thermal profiles of the ICM, specifically the density ($\rho$) and temperature (T), and the derived quantities which can be obtained through X-ray observations, to train the deep learning model. The radial profiles were defined on a logarithmic fixed grid of 48 points, centred on maximum of the density field, spanning from $0.02\, \textrm{R}_{500}$ to $2\, \textrm{R}_{500}$. We stress that in the analysis, the radius is not expressed as a function of $R_{500}$, even though we show the profiles with the normalised radius for clarity. 

\subsection{Dynamical state of the simulated sample}
\label{sec:sub1}
The clusters were categorised based on their intrinsic dynamical state, which was assessed as either relaxed or disturbed, using various estimators as detailed in \citet{Rasia2013}. The key intrinsic estimators employed were $f_{\rm s}=\textrm{M}_{\rm sub}/\textrm{M}_{500}$, representing the fraction of $ \textrm{M}_{500}$ accounted for substructures ($\textrm{M}_{\rm sub}$), and $\Delta_{\rm r}=|r_{\delta}-r_{\rm cm}|/\textrm{R}_{500}$, indicating the offset between the central density peak ($r_{\delta}$) and the center of mass ($r_{\rm cm}$) normalised to the aperture radius of $\textrm{R}_{500}$. 
To classify relaxed objects, we required that both $f_{\rm s}$ and $\Delta_{\rm r}$ were lower than 0.1, as indicated by previous studies \citep[e.g.][]{Cui2017,Cialone2018,DeLuca2021}.

These two dynamical parameters, $f_{\rm s}$ and $\Delta_{\rm r}$, were combined as in \cite{2023A&A...679A..51I}, resulting in the relaxation parameter $\chi$:
\begin{equation}
\chi=\frac{1}{2}\times \left(\frac{\Delta_{\rm r}-\Delta_{\rm r,med}}{|\Delta_{\rm r,quar}-\Delta_{\rm r,med}|} + \frac{f_{\rm s}-f_{\rm s,med}}{|f_{\rm s,quar}-f_{\rm s,med}|}\right).
\end{equation}
Here, $\Delta_{\rm r,med}$ and $f_{\rm s,med}$ represent the medians of the distributions of $\Delta_{\rm r}$ and $f_{\rm s}$, respectively, while $\Delta_{\rm r,quar}$ and $f_{\rm s,quar}$ denote the first or third quartiles, depending on whether the parameters of a specific cluster are smaller or larger than the median. According to this formulation, clusters with $\chi<0$ are categorised as relaxed, whereas those with $\chi > 0$ are classified as disturbed \citep{Rasia2013}. It is worth noting that, by construction, the relaxation parameter \( \chi \) tends to approximately divide the cluster sample into two equally-sized subsets of relaxed and unrelaxed systems. 
Since there is no clear expectation for the exact fraction of objects in one or the other class, nor an established consensus on the parameter or set of parameters or on the thresholds that would define a cluster as relaxed or disturbed, in this work we will focus on the most extreme objects and, thus, on the 30 or 50 clusters with the lowest and highest values of $\chi$. 
For further discussion, see, for example, \citet{Zhang2022} and \citet{Haggar2020}. Figure~\ref{fig1} in Appendix~\ref{app00} shows the temperature and density profiles of 100 randomly-drawn objects from \thethree\ sample.  Also shown are the 30 most relaxed and the 30 most disturbed clusters.

\subsection{Hydrostatic mass estimates}
\label{sec:sub2}
We derived the total 3D mass profiles using the hydrostatic equilibrium (HSE) assumption, which relates the 3D pressure gradient and density of the ICM to the gravitational mass enclosed within a given radius. To estimate the pressure gradients, parametric modelling of 3D ICM profiles may be inadequate for capturing the complexity of thermodynamic structures. For example, \cite{2021MNRAS.502.5115G} found that when applying parametric models to their simulated cluster sample, 50\% resulted in poor fits (chi-square value of > 10), which can result in inaccurate estimates of the derivatives. To address this, we adopt a model-agnostic, non-parametric framework that avoids restrictive assumptions concerning the underlying physical trends. In such approaches, 3D profiles are typically smoothed to suppress noise/irregularities while preserving genuine physical gradients. We use the algorithm  adapted from \cite{Cappellari2013b}, which implements one-dimensional locally linear weighted regression \citep{doi:10.1080/01621459.1979.10481038}\footnote{\url{https://pypi.org/project/loess/}} to compute the pressure derivatives. In this approach, 
we minimise the weighted least-squares error at each point $\textrm{R}_n$ as
\begin{equation}
\min_{\beta_0 , \beta_1} \sum_{n=1}^N \omega_n \left[ \textrm{P}_{n} - (\beta_0 + \beta_1 (\textrm{R}_{n}-\textrm{R}) ) \right]^2, \quad \big|\, \beta_1 \leq 0
\end{equation}
where $N$ is the number of points considered in the local neighbourhood, \( \textrm{P}_{n} \) is the pressure value at the \( n^{\text{th}} \) neighbourhood radial bin corresponding to the radial position \( \textrm{R}_{n} \), and \( \beta_0 \), \( \beta_1 \) (conditioned as $\beta_1 \leq 0$) are the best-fitting coefficients of the local linear model at radius $\textrm{R}$ representing the predicted best-fitting smooth pressure and pressure slope respectively. The weights \( \omega_n \) assigned to each data point are computed using a tri-cube weighting function as
\begin{equation}
\omega_n = \left( 1 - u_n^3 \right)^3, \quad u_n = \frac{| \textrm{R}_{n} - \textrm{R} |}{d}
\end{equation}
where \( u_n \) is the normalised distance of the data point \( \textrm{R}_n \) from the local radial point \(  \textrm{R}\), and \( d \) is the maximum distance considered for the local neighbourhood (i.e. $d={\rm R}_{N}-{\rm R}_{n}$). The smoothing parameter \( f \) controls the fraction of total data points included in the local regression neighbourhood (i.e. $N=f\times48$, where 48 is the total number of radial bins). It governs the smoothness of the fitted curve; smaller values of \( f \) emphasise local variations, while larger values lead to a smoother profile. Since, in our case, the profiles are logarithmically binned (with a higher density of points at lower radii), we consider 
$f$ as a dynamic parameter, which changes from 0.07 to 0.10 from outer to inner regions logarithmically. This ensures adaptive smoothing across different scales.

After performing the fit, we observed that a few points, particularly at small radii (where the profiles are relatively more irregular), resulted in best-fitting \( \beta_1 \) values of zero (the upper bound of the fit). This suggests that at these points, the fit actually favours positive values of \( \beta_1 \), which is inconsistent with the hydrostatic equilibrium assumption. To address this, we undertook the fit again after removing these problematic points.
\begin{figure}
		\includegraphics[width=0.45\textwidth]{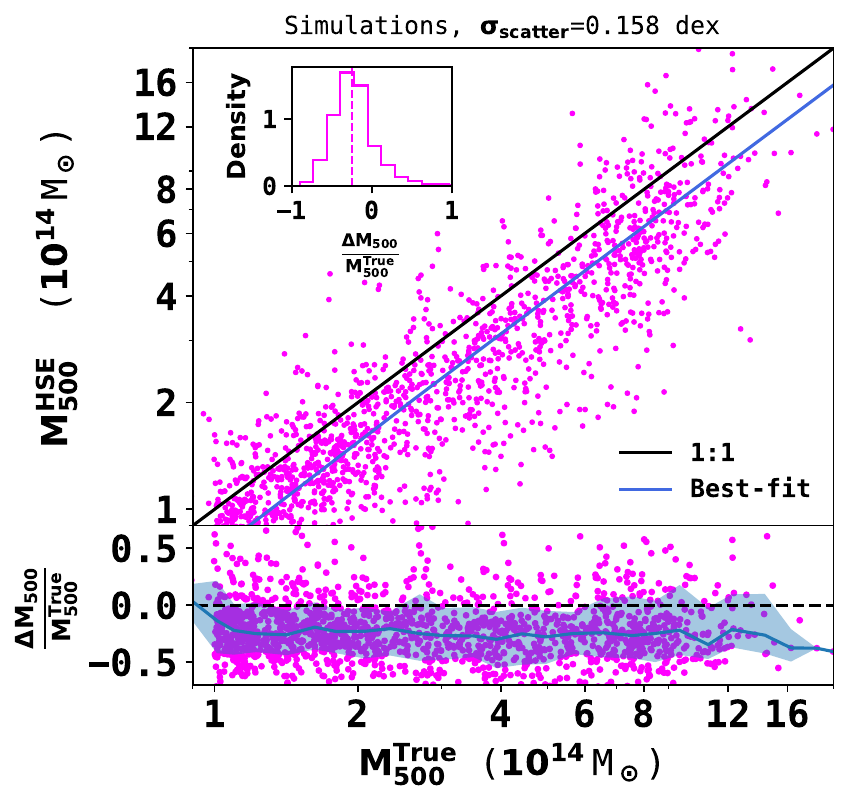}
		\caption{\footnotesize Comparison of the hydrostatic mass estimates \( \text{M}^{\text{HSE}}_{500} \) to the true mass \( \text{M}^{\text{True}}_{500} \) for a sample of 1655 galaxy clusters in \thethree. The blue line indicates the best-fitting linear relation, and the black line corresponds to the  1-to-1  relation. The blue line and shaded region in the lower panel show the median and 1$\sigma$ dispersion of the fractional residual distribution, respectively, considering logarithmic binning. The inset plot shows the distribution of fractional residuals and the vertical dashed lines show the median of the fractional dispersion of $-0.24$.
        }
		\label{fig2}
\end{figure}
\begin{figure*}
\centering
		\includegraphics[width=.9\textwidth]{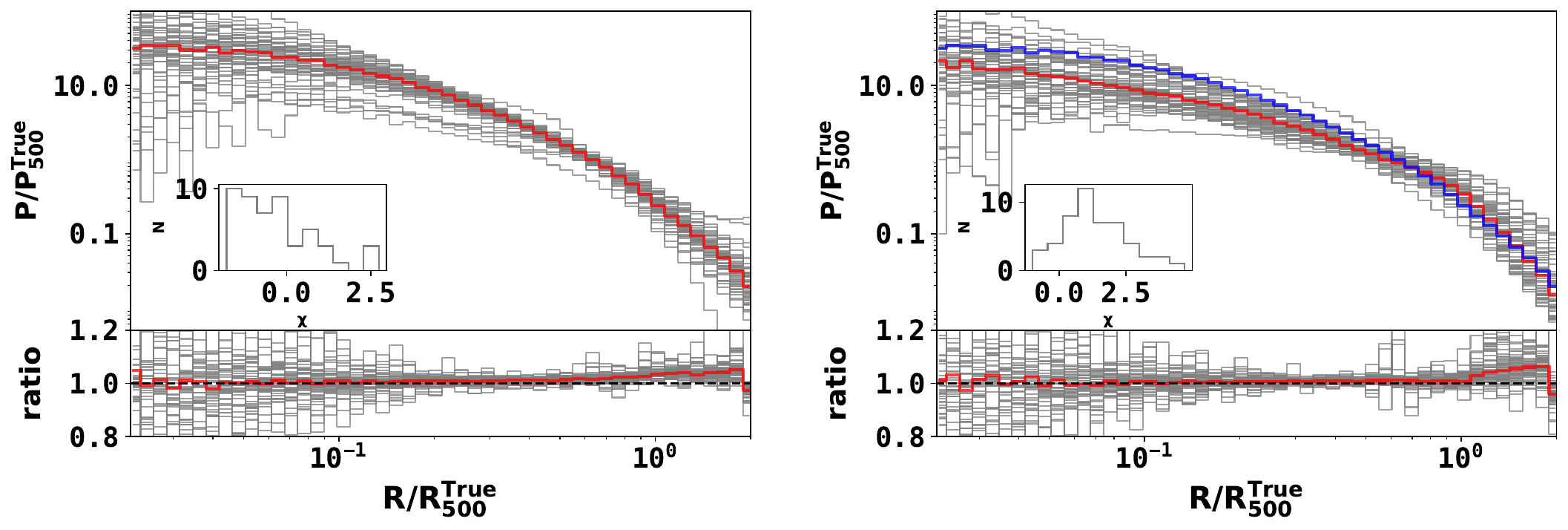}
		\caption{\footnotesize The grey lines show the pressure profiles of the 50 clusters having the lowest (left panel) and highest (right panel) fractional dispersions. The thick red lines indicate the corresponding median profiles. The bottom panel displays the ratio of reconstructed to true pressure profiles. On the right panel, we have also plotted the median profile for the lowest dispersion set with a thick blue line for comparison. The inset plot illustrates the distribution of the relaxation parameter $\chi$ for the same clusters.	}
		\label{fig3a}
\end{figure*}
Given the smooth 3D pressure derivatives from the above regression and 3D density profiles, the hydrostatic mass within a radial distance R from the center of the clusters is then computed as
\begin{equation}
{\textrm M}(<\textrm{R}) =- \frac{r^2}{G} \frac{d\textrm{P}}{d\textrm{R}} \frac{1}{\rho(\textrm{R})}.
\label{H:mass}
\end{equation}
Finally, we extracted the hydrostatic mass within  $\textrm{R}^{\textrm{HSE}}_{500}$, (i.e $\textrm{M}^{\textrm{HSE}}_{500}$) using these derived mass profiles. In Fig.~\ref{fig2}, we show the relationship between the hydrostatic mass \( \textrm{M}^{\text{HSE}}_{500} \) and the true mass \( \textrm{M}^{\text{True}}_{500} \) with a best-fitting line given by 
\begin{equation}
{\textrm M}^{\text{HSE}}_{500} = ( 0.78 \pm 0.01) \, {\textrm M}^{\text{True}}_{500} - (0.02 \pm 0.07) \times 10^{14}{\textrm M}_\odot.
\end{equation}
We find a median fractional residual at R$_{500}$,  $\frac{\Delta \textrm{M}_{500}}{\textrm{M}_{500}^{\textrm{True}}}=\frac{\textrm{M}_{500}^{\textrm{HSE}}-\textrm{M}_{500}^{\textrm{True}}}{\textrm{M}_{500}^{\textrm{True}}}$, and its associated 1$\sigma$ dispersion (16th–84th percentile range) to be equal to $-0.24\pm0.23$.This corresponds to an average hydrostatic bias of $(1 - b) = 0.75 \pm 0.23$, consistent with previous results by \citet{2020A&A...634A.113A} and \citet{2023MNRAS.518.4238G}, both of which were also derived using data from \thethree.  The scatter between the true mass and the hydrostatic mass is $\sigma_{\rm scatter}=0.158$ dex, reflecting the intrinsic variation beyond the mean bias.
\begin{figure}
		\includegraphics[width=0.45\textwidth]{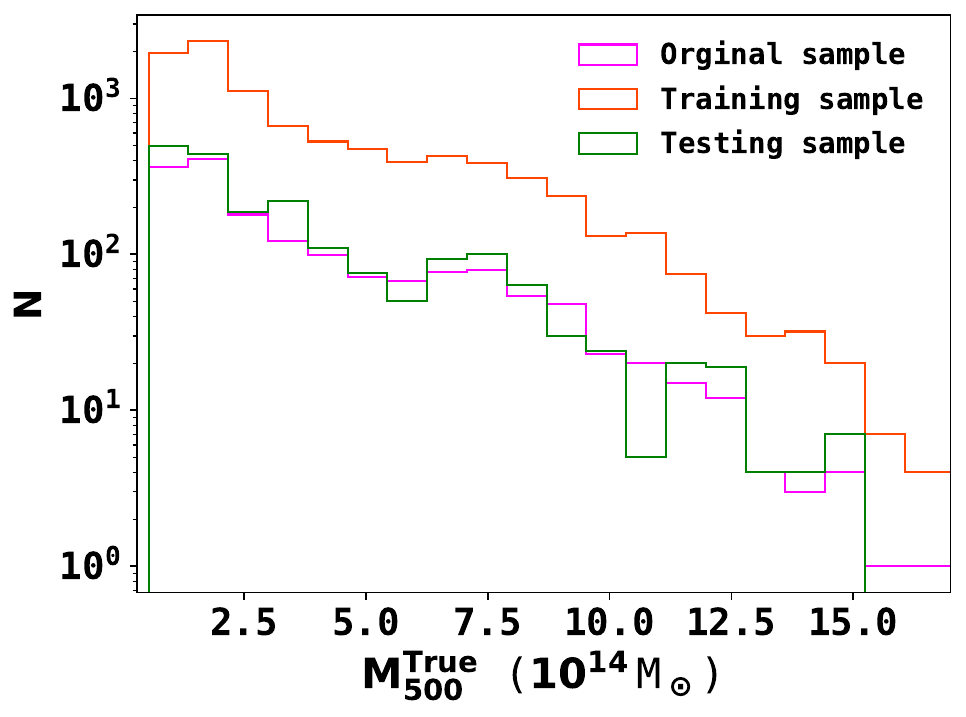}
		\caption{\footnotesize The number distribution of galaxy clusters in the \thethree, shown as a function of \( \text{M}^{\text{True}}_{500} \). The magenta histogram represents the original sample (1655 clusters), while the orange and green histograms represent the augmented training sample (9305 clusters built with 1305 clusters) and testing (1950 clusters built with 350 clusters) sample, respectively.
		}
		\label{fig3}
\end{figure}

To better understand the dispersion, we present in Fig.~\ref{fig3a} the reconstructed pressure profiles obtained using local polynomial fitting for the 50 clusters with the lowest (left-panel) and highest (right-panel) mass bias, corresponding to the sets with the lowest and highest absolute value (modulus) of dispersion, respectively. These profiles are scaled by self-similar parameter ${\rm P}_{500}^{\mathrm{True}} = 1.65 \times 10^{-3} \, E(z)^{8/3} \left( \frac{{\rm M}_{500}^{\mathrm{True}}}{3 \times 10^{14} \, M_\odot} \right)^{2/3} \, \mathrm{keV \, cm^{-3}}$.
 For the least biased cluster set, the ratio between the true and reconstructed profiles remains very low, with deviations typically below 5\% around R$_{500}$. Similarly, the dispersion in the highest biased set increases beyond R$_{500}$ but remains mostly below 10\% around R$_{500}$.   The inset figures further reveal that the regular clusters predominantly belong to the relaxed category ($\chi < 0$), whereas the outlier clusters are mainly disturbed objects ($\chi > 0$). These results suggest that the large discrepancies in mass estimates for the outlier clusters can be attributed to their intrinsic dynamical state. Furthermore, in the central region and up to 0.6\,R$_{500}$, we find that the median pressure profile of the most biased set is approximately half that of the least biased set. This suggests that the highly biased clusters are predominantly composed of disturbed, non-cool-core systems, whereas the low-bias clusters are primarily relaxed cool-core clusters.

\subsection{Training and testing samples}
\label{twopointfour}
To enhance the dataset and improve model training, we augmented the sample by randomly placing simulated galaxy clusters at various redshifts within the range \( 0 \leq z \leq 1.2 \). For each placement, we recomputed 
the corresponding \( \textrm{M}_{500} \), accounting for its dependence on redshift through the critical density. Specifically, clusters originally at redshifts \( z = 0.07 \), \( 0.33 \) and \( 0.95 \) were randomly reassigned 
to the redshift bins \([0.00 - 0.16]\), \([0.16 - 0.65]\) and \([0.65 - 1.2]\), respectively. These bins were chosen to preserve any potential redshift-dependent trends present in the original cluster samples, while allowing for variability within each bin. 
This approach enabled the same cluster to appear at different redshifts, thus incorporating redshift-dependent effects into the augmented sample. Furthermore, to explore the impact of radial coverage and binning -- factors relevant 
to observational limitations \citep[e.g.][]{2024A&A...688A.219C}-- we also modified the high-resolution simulated profiles, originally defined at 48 grid points across \([0.02 - 2]\, \textrm{R}_{500}\). The number of radial bins was reduced (with a minimum of 7), 
and both the inner and outer radial boundaries were randomly varied to allow profiles of different radial ranges. The first radial bin was selected within \([0.02 - 0.3]\, \textrm{R}_{500}\) and the last bin within \([0.75 - 2]\, \textrm{R}_{500}\), 
resulting in a diverse set of profiles that more realistically reflect observational data quality. Importantly, care was taken to ensure that training and testing samples remained disjoint throughout the augmentation process. 
The original dataset was first divided into 1305 training and 350 testing clusters and each subset was then independently expanded to 9305 and 1950 clusters, respectively. Figure~\ref{fig3} shows the distribution of cluster masses across the different samples.

\section{Graph Neural Networks}
In this Section, we present an overview of Graph Neural Networks (GNNs). GNNs have already been applied to different astrophysical and cosmological studies \citep{2022OJAp....5E..18M,2021arXiv210609761C, 2022ApJ...935...30V,2022mla..confE..19J,2022ApJ...937..115V,2022ApJ...941....7J, 2023arXiv230612327W,2023MLS&T...4d5002L,2024ApJ...976...37W,2024ApJ...965..101C}. GNNs offer several advantages over CNNs, particularly when dealing with irregular data. While CNNs rely on predefined grid structures, GNNs can handle data of varying sizes and resolutions, making them well-suited for profiles or maps with different levels of granularity. 


\subsection{General Concepts on Graphs}
A graph is represented as a combination of a set of nodes, \( V \), and a set of edges, \( E \), connecting these nodes: \( G = (V, E) \). In the context of GNNs, each node \( i \in V \) is characterised by a feature vector that encodes relevant information specific to that node. The edges \( (i, j) \in E \) indicate relationships between the nodes \( i \) and \( j \). The structure of the graph is captured by an adjacency matrix \( A \), where \( A_{ij} = 1 \) if nodes \( i \) and \( j \) are connected by an edge, and \( A_{ij} = 0 \) otherwise. The properties of the adjacency matrix depend on whether the graph is undirected or directed. In an undirected graph, the edge between nodes indicates a bidirectional connection, which results in a symmetric adjacency matrix \( A_{ij} = A_{ji} \). In cases where there are no reciprocal directional relationships in the nodes, $A_{ij}$ is not symmetric. Since each node is connected exclusively to its neighbouring nodes, the neighbourhood of a node \( i \), denoted \( \mathcal{N}(i) \), consists of all nodes \( j \) that are directly connected to node \( i \). Formally, this is expressed as
\begin{equation}
\mathcal{N}(i) = \{ j \in V : (i, j) \in E \}.
\end{equation}

\subsection{Graphs from ICM profiles}
In our approach, the radial profiles of the ICM are represented as graphs, where each node corresponds to a radial grid point, and the edges capture the physical relationships between neighbouring points, such as spatial proximity. 
Each node in the graph is assigned features representing the 3D physical properties of the ICM: 
temperature, density, pressure, total gas mass, and the corresponding radius. 
In this approach, \( \mathbf{x}_i \) represents the feature vector of the grid point \( i \), where \( \mathbf{x}_i \in \mathbb{R}^5 \) contains the following five features: radius ($\textrm{R}_i$), density ($\rho_i$), temperature ($\textrm{T}_i$), pressure ($\textrm{P}_i$) and gas mass ($\textrm{M}_{g,i}$) as
\begin{equation}
\mathbf{x}_i = \begin{bmatrix}
\textrm{R}_i,  \rho_i, \textrm{T}_i,\textrm{P}_i, \textrm{M}_{g,i}\end{bmatrix}
\end{equation}
where R is in physical units and not in units of R$_{500}$.
Including derived pressure and gas mass as input features alongside the density and temperature could enhance the model's robustness by providing additional physical context, allowing the neural network to learn relationships without predefined equations. For instance, the gas mass, as a cumulative property, will help to capture large scale spatial information, stabilising local variations and offering a global perspective on the ICM structure.
We represent the complete set of feature vectors of all nodes of a given cluster as a matrix \( \mathbf{X} \), where \( \mathbf{X} \in \mathbb{R}^{n \times 5} \). Here, \( n \) represents the total number of nodes in the graph, which vary between $7-48$, corresponding to the number of radial grid points and each row of the  \( \mathbf{X} \) corresponds to the feature vector of a specific node, thereby capturing the physical properties of the ICM across the entire grid.  The set of edges \( E \) is constructed based on the spatial relationships between grid points. Specifically, in our model, each node \( i \) is assumed to be connected with every other node  \( j \)\footnote{We also considered as case where each node \( i \) is connected only to its immediate neighbours only, \( i-1 \) and \( i+1 \). This resulted in about a 2\%-4\% increase in the scatter of the recovered masses.}. 
Furthermore, nodes are assumed to be connected symmetrically, so the adjacency matrix satisfies, \( A_{ij} = A_{ji} \), enabling bidirectional information flow. The graph-based representations are then processed by the GNN described 
in the following subsections. 

\subsection{Message Passing in GNNs}
Unlike traditional neural networks, which are tailored for structured data such as images (grids) or sequences (ordered tokens), GNNs are designed to handle graph-structured data, where the underlying topology can be irregular and non-Euclidean.
The key operations in GNNs are neighbourhood aggregation and feature update, where a node's feature vector is aggregated from the features of its neighbouring nodes and then updated through a learnable transformation.
The forward propagation in a GNN layer can be generally described using the message-passing framework as
\begin{equation}
\mathbf{h}_v^{(k+1)} = \phi \left( \mathbf{h}_v^{(k)}, \bigoplus_{u \in \mathcal{N}(v)} \psi \left( \mathbf{h}_v^{(k)}, \mathbf{h}_u^{(k)} \right) \right)
\end{equation}
where $\mathbf{h}_i^{(k)}$ is node $i$'s feature vector at layer $k$ and $\mathcal{N}(i)$ denotes its neighbors. The message function $\psi$ computes interactions between nodes, $\bigoplus$ aggregates neighbor messages (using sum, mean, max, or attention), and $\phi$ updates the node state (typically a neural network with non-linearity).

For instance, Graph Convolutional Networks (GCNs) \citep{2016arXiv160902907K} update the node features using a sum of normalised neighbourhood embeddings, by combining each node's feature with those of its neighbours after normalising them by their degrees, and thus effectively applying convolutional filters on the graph. Graph Attention Networks (GATs) \citep{2017arXiv171010903V}, on the other hand, employ attention mechanisms to assign varying weights to neighbouring nodes, capturing their differing influences through learnable attention coefficients. While GCN or GAT typically operate in a transductive setting -- learning node embeddings based on the fixed graph structure seen during training -- GraphSAGE networks \citep{2017arXiv170602216H} introduces an explicitly inductive approach by learning parameterised aggregation functions (e.g., mean, LSTM, pooling) that can be applied to the features of sampled neighbors, allowing it to generate embeddings for unseen nodes or entirely new graphs at inference time. Similarly, Graph Transformer networks \citep{2020arXiv200903509S} extend the inductive capabilities by incorporating multi-head self-attention mechanisms over a node neighbors, enabling more expressive and flexible neighborhood aggregation that can generalise to unseen nodes or graphs. Another approach is that of the Gated Graph Neural Networks (GGNNs) \citep{2015arXiv151105493L} that incorporate recurrent units such as Gated Recurrent Units (GRUs) to iteratively update the node states, enabling the capture of more complex and long-term dependencies within the graph. Each of these models showcases a different strategy for refining the node representations based on their neighbourhood information, enabling GNNs to extract intricate patterns from graph-structured data.

\subsection{Model Architecture}
We implement our graph neural network (GNN) architecture using the \texttt{PyTorch Geometric} library, which provides a wide range of message-passing layers. We consider several types of of message-passing layers: Graph Convolutional Network ({\tt GCNConv}), Graph Attention Network ({\tt GATConv}), GraphSAGE ({\tt SAGEConv}), and Graph Transformer ({\tt TransformerConv}). As discussed in the previous Section, these layers enable the network to aggregate information from neighboring nodes based on different mechanisms, allowing it to capture both local and global structural patterns within the graph. In the current implementation, we do not use multi-head attention for the Transformer GNN; the layer is applied with a single attention head to maintain a fixed embedding size and simplicity.

Following the message-passing layers, we apply one of several global pooling operations to aggregate the node-level features into a fixed-size graph-level representation. Specifically, we consider {\tt global mean pooling}, {\tt global sum pooling}, {\tt global max pooling} and an {\tt attention-based pooling} strategy. In the attention-based variant, learnable attention weights are used to highlight the most informative nodes during the pooling operation. Specifically, a linear layer computes a score for each node, which is then normalised and used to weight the node features during aggregation, allowing the network to focus on key nodes that contribute most to the overall graph representation.

The first message-passing layer receives node features of size 5, which correspond to the temperature, density, pressure, gas mass of the ICM, and the radius associated with each node. These features are projected into a higher-dimensional embedding space. Subsequent message-passing layers refine these embeddings while maintaining the same dimensionality, enabling the network to learn rich, hierarchical representations of the graph. Each layer is followed by batch normalisation and a ReLU activation function to stabilise training and introduce non-linearity, improving the model's capacity to capture complex relationships between node features.

To incorporate external redshift information, we concatenate the redshift value to the pooled graph embedding, forming a combined representation that captures both structural and redshift-dependent information. This combined embedding is then passed through a series of fully connected linear layers. These layers are followed by batch normalisation and ReLU activation. The final linear layer outputs a single scalar prediction, representing the cluster mass ${\rm M}_{500}$, which is the target variable of the model.

\subsection{Model training}
Before training the model with the sample described in Sect.~\ref{twopointfour}, we first normalised the pressure, gas mass, radius, and total mass respectively by $10^{-10} \, \textrm{erg} \, \textrm{cm}^{-3}$, $10^{14}\textrm{M}_{\odot}$, $10^{3} \, \textrm{kpc}$, and $10^{15} \, \textrm{M}_{\odot}$, so that their maximum value is close to unity. The ICM density (in $\textrm{cm}^{-3}$) and temperature (in KeV) profiles were not normalised since their values are already within a manageable scale and do not present issues of large magnitude that could destabilise gradient descent during training. Thereafter, all quantities were transformed to base-10 logarithmic space to further stabilise the training and better capture multiplicative relationships between variables.
We employed the mean absolute error (MAE) as a loss function, defined as
\begin{equation}
\textrm{MAE} = \frac{1}{N} \sum_{l=1}^{N} |\log\textrm{M}_{500}^l - \log \hat{\textrm{M}}_{500}^l|
\end{equation}
where \(\textrm{M}_{500}^l\) and \(\hat{\textrm{M}}_{500}^l\) are the true and predicted values of 3D \(\textrm{M}_{500}\) for the galaxy cluster \(l\), and $N$ is the number of clusters in the sample (training or testing set). 

The optimal values of hyper-parameters discussed in the next paragraph were obtained using {\tt Optuna}\footnote{\url{https://optuna.org/}} \citep{akiba2019optuna}, which efficiently searches for the best combinations by minimising the average test loss utilising the Tree-structured Parzen Estimator \citep{2023arXiv230411127W}.

The Adam optimiser was used with an initial learning rate of \(9.9\times10^{-4}\) and weight decay set to \(2\times10^{-3}\) to mitigate overfitting. We used the {\tt ReduceLROnPlateau} scheduler, which adaptively lowers the learning rate when validation loss plateaus to improve convergence. The reduction scale was controlled by the {\tt lr factor} of $0.5$, with smaller values enabling finer adjustments. The training set was divided into a batch size of \(128\) graphs. Gradients were calculated through backpropagation, and the optimiser updates the model parameters. After each epoch, the model was evaluated on the training and testing sets to assess performance. At round 500 iterations, we find the value of the loss function is well flattened, indicating minimal further improvement. At this point, we considered the training to have sufficiently converged and stopped the process. We used three \texttt{TransformerConv} layers with an embedding size of 128 along with global max pooling. The graph-level embedding was followed by three fully connected linear layers: the first reduced the dimensionality from 129 to 60, the second from 60 to 30, and the final output layer produced a scalar prediction.

 Table~~\ref{tab:hyperparameters} in Appendix~\ref{app000} lists the hyper-parameters used in this work. Table~\ref{tab:gcn_architecture} in  Appendix~\ref{app000} presents a comprehensive overview of our GNN architecture, detailing the type, input size, output size, and function of each layer.

\begin{figure}
		\includegraphics[width=0.45\textwidth]{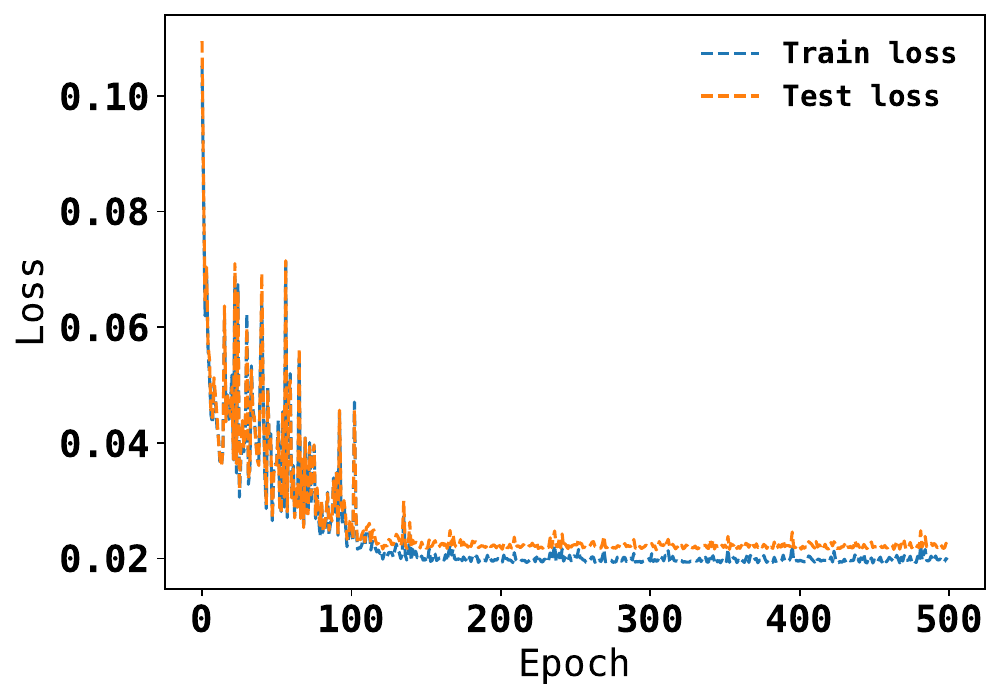}
        \centering
		\caption{\footnotesize Train and test loss in the evaluation mode over epochs for one of the ensemble models. Both train and test losses indicate consistent learning, with the final loss at around 0.02 for the training and testing sets.
		}
		\label{fig4}
\end{figure}

To account for the effects of random initialisation, we trained a total of 50 individual models, each initialised with different random weights. By aggregating the predictions from these models, we aim to mitigate the risk of any specific initialisation, thereby improving the overall robustness and generalisation capability of our ensemble. Each model was weighted according to its validation loss, allowing us to give greater importance to models that perform better. The final predicted mass, given the input {\bf X}, is defined by the weighted mean over the ensemble.
The final loss values showed minimal variation across different models. In Fig.~\ref {fig4}, we plot the training and testing losses for one of the models. The gradual decline in both losses indicates stable learning, converging to final values.

\begin{figure*}
    \centering
    \begin{minipage}{0.45\linewidth}
        \includegraphics[width=\textwidth]{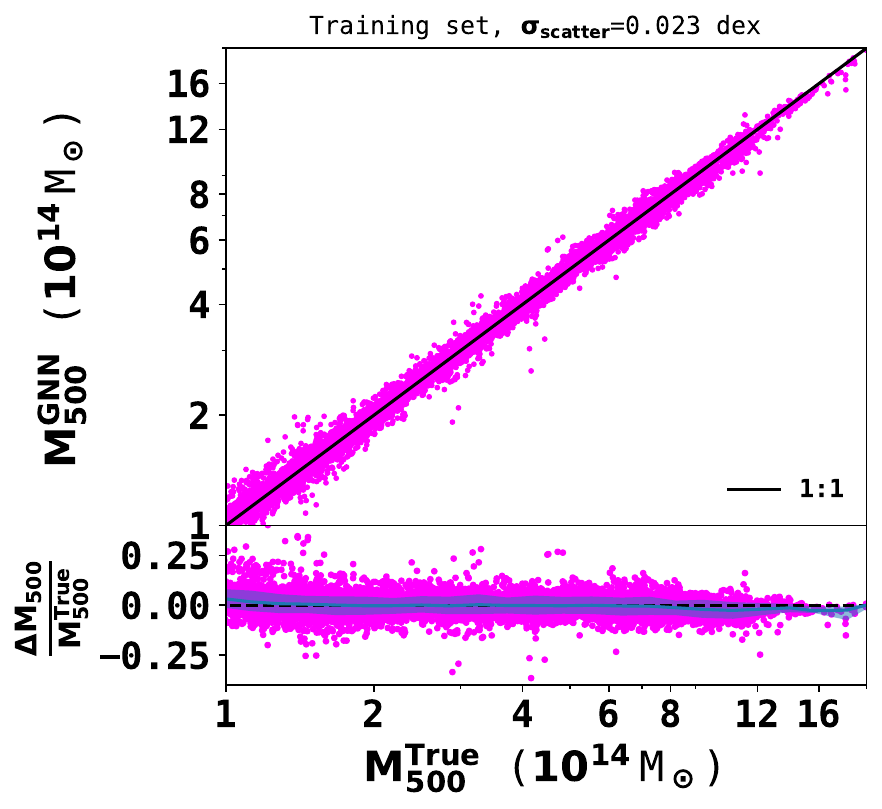} %
    \end{minipage}
    \begin{minipage}{0.45\linewidth}
        \includegraphics[width=\textwidth]{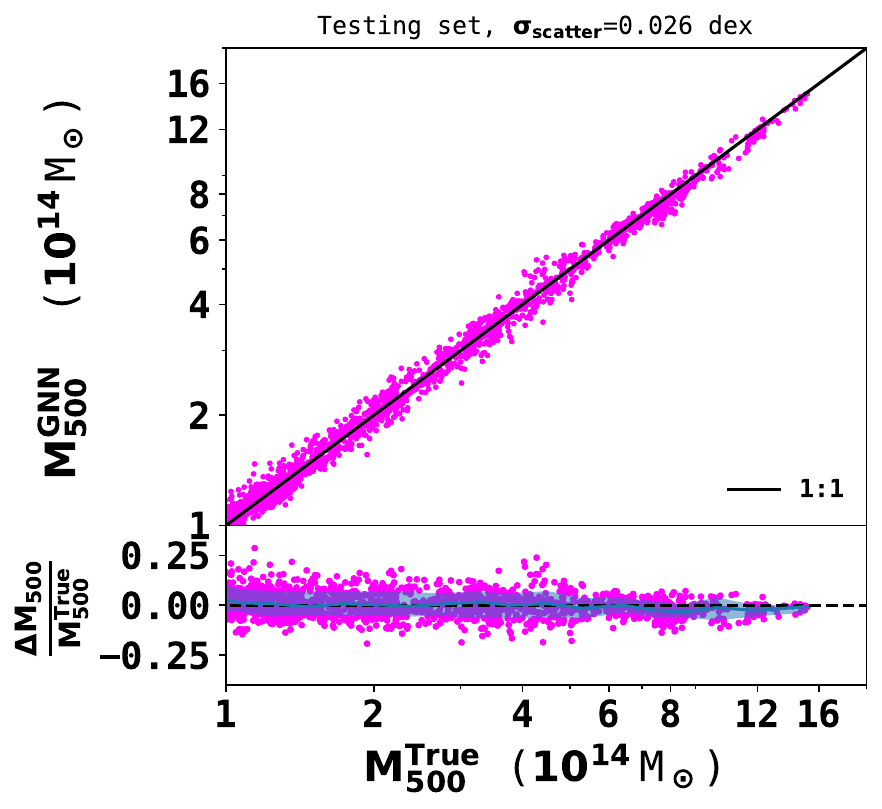}
    \end{minipage}
    \caption{\footnotesize Distribution of GNN predicted masses, $\text{M}^{\text{GNN}}_{500}$, versus true cluster masses, $\text{M}^{\text{True}}_{500}$, for the training (left) and testing (right) samples of sizes  9305 and 1950 respectively. The black line represents a  1-to-1  relation. The bottom panel in both panels shows the fractional residual distribution with a solid blue line and shaded regions representing the median and 1$\sigma$ dispersion, respectively. We find a 1$\sigma$ dispersion of about 7\%  and 6\% for the testing and training samples respectively}. 
    \label{fig5}
\end{figure*}

\section{Model Evaluation}
In this Section, we evaluate the performance of our model in predicting galaxy cluster masses, specifically the characteristic mass M$_{500}$. We examine the model’s robustness to observational constraints, analyse its sensitivity to cluster dynamical states, and compare its performance to the traditional hydrostatic mass estimation method. Finally, we use interpretability tools to understand the contributions of various radial thermal profile features to the model’s predictions.

\subsection{Performance of the GNN model}
To evaluate the generalisability of our GNN model, we tested it on a reserved dataset spanning radial profiles which have an extent up to [0.75-2] R$_{500}$. This dataset covers an extensive range of ICM properties across radii, allowing us to rigorously assess the model's capacity to perform accurately on unseen data. Figure~\ref{fig5} displays the distribution of GNN predicted masses, $\text{M}^{\text{GNN}}_{500}$,  versus true cluster masses, $\text{M}^{\text{True}}_{500}$, for both the testing (right panel) and training (left panel) samples, illustrating the model's ability to align its predictions closely with true values. This strong alignment confirms that the GNN effectively captures the patterns of the mass distribution across various scales, demonstrating robust generalisability beyond the training set. In particular, we observe that the median fractional residuals, 
$\Delta \textrm{M}_{500}/\textrm{M}_{500}^{\textrm{True}}=(\textrm{M}_{500}^{\textrm{GNN}}-\textrm{M}_{500}^{\textrm{True}})/\textrm{M}_{500}^{\textrm{True}}$, remain near zero throughout the mass range. We find the fractional residual to be $ 0.00\pm0.06$ and $0.00\pm0.05$ for the testing and training samples, respectively. Similarly, the mass bias, $(1 - b)$, is equal to $1.00 \pm 0.06$ for the testing sample and $1.00 \pm 0.05$ for the training sample.  Finally, we find the scatter between the true and predicted mass to be  $0.026$ dex and  $0.023$ dex for testing and testing samples, respectively.   In terms of mass estimation accuracy, the model achieves an R$^2$ score of 0.99 on both testing and training samples, reflecting its high fidelity in reproducing true masses with minimal error. This level of accuracy suggests that the GNN can reliably estimate masses across diverse radial profiles, underlining its potential for applications requiring accurate mass distribution predictions. 

To our knowledge, this is the first model to attain sub-7\% scatter using only low-dimensional spherically symmetric radial profiles as input, providing a robust and efficient alternative to image-based methods. \citet{2019ApJ...884...33G} employed ensemble regressors on morphological features extracted from simulated X-ray maps and reported a mass scatter of $\sim$16\%. Similarly, \citet{2019ApJ...876...82N} and \citet{2020MNRAS.499.3445Y} achieved scatters of $\sim$12\% and $\sim$16\%, respectively, using neural networks trained on Chandra-like mocks generated from simulations. On the other hand, \citet{2022NatAs...6.1325D} demonstrated that CNN-based estimators using SZ observations can achieve $\sim$10\% scatter which is comparable to the accuracy achieved in our work using X-ray data. \citet{2024A&A...682A.132K} applied CNNs to multi-band eROSITA mocks and reported a scatter of $\sim$19\%, meanwhile, \citet{2023MNRAS.524.3289H} using similar architecture achieved a best-case scatter of about 16\%. Despite using richer, higher-dimensional inputs, both studies report higher scatter than our GNN model trained solely on single-band radial profiles, which attains substantially lower scatter by utilising lower-dimensional, physically interpretable features. 

While direct, one-to-one comparisons between these studies remain inherently difficult due to differences in simulation suites, input representations, model architectures, training objectives, and sample selection, our comparisons should be viewed as indicative rather than definitive. A standardised benchmark using consistent simulation and observation pipelines will be crucial for a robust comparison of future machine learning-based mass estimators and will be performed in the future studies.

To assess the impact of core regions on mass estimation, we also trained a GNN model on core-excised profiles, as discussed in Appendix~\ref{sec:Core-excised GCN modeling}. By excluding the AGN-dominated central ICM (within $\sim0.1,\textrm{R}_{500}$), we observe a slight improvement in the scatter of predicted masses (0.025 dex in testing sample). In Appendix~\ref{app0}, we trained two additional variants of our fiducial model. The first variant used only three features -- radius, density, and temperature -- while the second included six features, incorporating entropy in addition to those used in the fiducial model. For the three feature model, the bias and fractional residuals are unbiased, but the scatter in the testing sample is about 1.5 times larger than for the fiducial model. For the six-feature model, we observe no reduction in scatter compared to our fiducial model. Moreover, the inclusion of entropy as a feature does not necessarily translate to more physically robust predictions. From an observational standpoint, entropy measurements are known to be significantly affected by gas clumping, which can introduce systematic uncertainties.

Similarly, in Appendix~\ref{app1}, we investigated the impact of a redshift dependence on model training by using distinct redshift bins for the training and testing samples. We find that when the model is trained on low ($z=0.06$) and intermediate ($z=0.33$) redshift samples -- both augmented within the redshift range $0 \leq z \leq 1.2$ (with random binning sizes as done for the main case) -- and then tested on a high-redshift sample at $z=0.96$ (also augmented over the same redshift range and binning).  We find that the fractional dispersion for the testing sample has a median of $-0.04 \pm 0.07$, with a median bias of $0.96 \pm 0.08$ and a scatter of 0.061 dex. In contrast, for the training sample, the fractional dispersion and bias are $0.00 \pm 0.05$ and $1.00 \pm 0.06$, respectively, with a scatter of 0.025 dex. The larger dispersion and bias on the testing sample compared to the training sample suggest that the model may be overfitting to the lower redshift training data.
This comparison illustrates that the redshift distribution of the training data plays an important role. In particular, the underrepresentation of high-$z$ clusters in the training set limits the model’s ability to capture the evolving thermodynamic and structural properties of the ICM, leading to a mild systematic bias and increased scatter in the predictions.

\begin{figure}
		\includegraphics[width=0.45\textwidth]{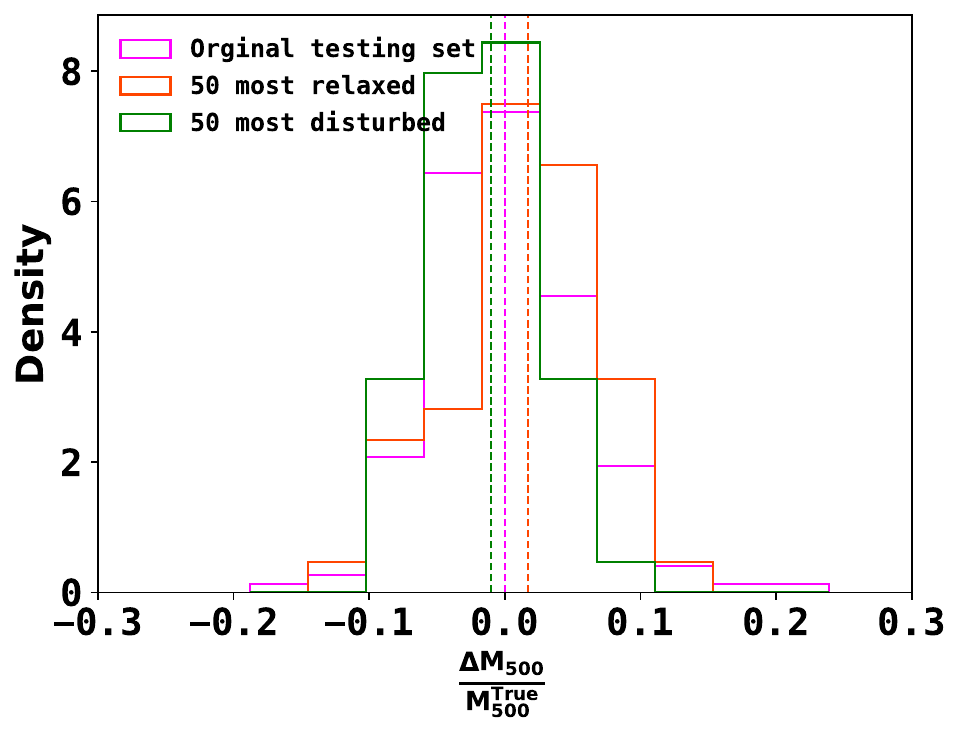}
                \centering

		\caption{\footnotesize Histogram of residual distributions for the entire unaugmented test sample of 350 galaxy clusters, as well as for the 50 most relaxed and 50 most disturbed clusters within this sample. Vertical dashed lines show the corresponding median of distributions. We find comparable performance of the GNN model across different dynamical states.
		}
		\label{fig6}
\end{figure}
\subsection{Impact of cluster dynamical state on model performance}
To assess how the dynamical state influences the model performance, we evaluated the GNN separately on relaxed and disturbed clusters in the original, unaugmented testing sample of 350 clusters. Figure~\ref {fig6} presents the histogram of residual distributions for the entire unaugmented test sample, as well as for the 50 most relaxed and 50 most disturbed clusters within this sample. Our results indicate that the residuals are comparable across these sub-samples. Across the test set, the median residuals (with 1$\sigma$ dispersions) are $0.00 \pm 0.05$ (scatter $0.023$ dex), $0.02 \pm 0.05$ (scatter $0.017$ dex) and $-0.01 \pm 0.04$ (scatter $0.023$ dex) for the unaugmented high-resolution test sample of 350 clusters, the 50 most relaxed clusters and the 50 most disturbed clusters, respectively. In all cases, the scatter is less than $7\%$, demonstrating that the GNN model maintains consistent performance for both relaxed and disturbed morphologies. 
\begin{figure}
		\includegraphics[width=0.45\textwidth]{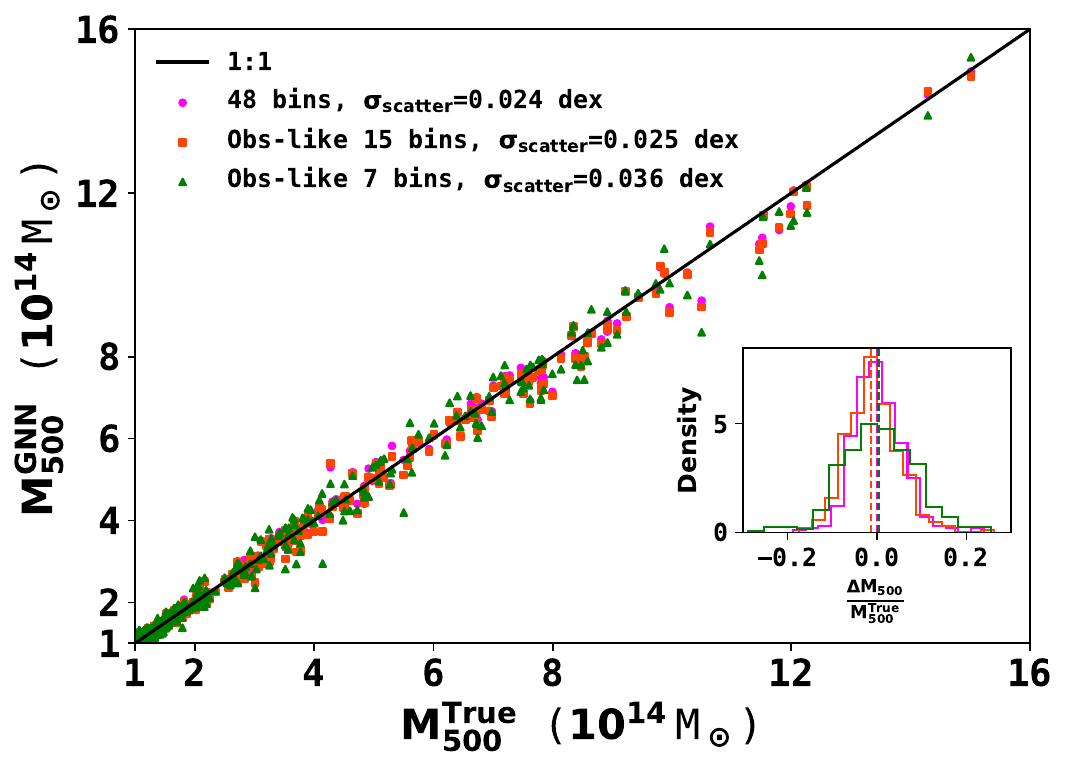}
                \centering

		\caption{\footnotesize Comparison of predicted and true galaxy cluster masses for an unaugmented test sample of 350 clusters. 
        The black line represents a  1-to-1  relation. Magenta, orange and green points represent predictions, respectively, using ICM profiles truncated at 2R$_{500}$, R$_{500}$ and 0.75 R$_{500}$, with 48, 15 and 7 bins. 
        The insert plot shows the corresponding fractional residuals with the vertical lines representing the medians. 
		}
		\label{fig7}
\end{figure}
\subsection{Performance with respect to observational-like profiles}
\label{sub_A1}
To simulate observational constraints, we truncated the ICM profiles to R$_{500}$ and 0.75 R$_{500}$, assigning 15 and 7 observational bins, respectively. These truncations mimic the radial limits of current X-ray observations, approximating the depth of XMM-Newton's deep and shallow observations. This approach tests the GNN's robustness when spatial information is limited. Despite the reduced radial coverage, the GNN achieves a robust mass estimation accuracy with a precision of better than 10\% on average, with respect to the true mass. Figure~\ref {fig7} compares predicted and true cluster masses in the observational-like profile datasets, using the original testing sample of 350 galaxy clusters. The model captures the overall mass distribution well, even with restricted radial data. The median fractional residual error and the associated 1$\sigma$ dispersion are consistent with zero, with values of $0.00\pm0.05$ (scatter 0.024 dex),  $0.01 \pm 0.06$ (scatter $0.025$ dex) and $0.00 \pm 0.08$ (scatter $0.036$ dex) for the 48 (fine bining), 15 bin and 7 bin cases, respectively. These results highlight the model's adaptability to incomplete data while maintaining strong predictive performance. In Table~\ref{tab:fitA}, we give the best-fit linear relation between the GNN derived masses versus true masses. Finally, we tested a model in which the central region (< 0.1 R$_{500}$) of the input profiles was excluded from the mass estimation using our fiducial model. The model performance remains very similar across all three cases -- fine binning and coarse binning with 15 and 7 bins -- with scatters of 0.024, 0.025 and 0.027 dex, respectively. This performance is similar to the results obtained with the core-excised model where the model was trained with input profiles having radii > 0.1 R$_{500}$. This reaffirms that feedback processes, which dominate in the central region, diminish its constraining power, thereby limiting the impact of its exclusion on the model’s accuracy.

\begin{table}
\resizebox{\columnwidth}{!} {
    \centering
    \begin{threeparttable}
    \begin{tabular}{ccc}
        \toprule
        \textbf{Case} & \multicolumn{2}{c}{${\textrm M}^{\text{GNN}}_{500} = {\rm A} \times {\textrm M}^{\text{True}}_{500} + {\rm B} \times 10^{14} \, {\textrm M}_\odot$} \\
        \cmidrule(lr){2-3}
        & A & B \\
        \midrule
        Fine bins & $1.016\pm0.004$ & $-0.043\pm0.020$ \\
        Observational like 15 bins & $1.020\pm0.004$ & $-0.021\pm0.020$ \\
        Observational like 7 bins & $1.023\pm0.009$ & $-0.056\pm0.027$ \\
        \bottomrule
    \end{tabular}
    \caption{\footnotesize Best-fit relation between the GNN and true masses for different cases considered in Sec.~\ref{sub_A1}.}
    \label{tab:fitA}
    \end{threeparttable}
    }
\end{table}

\begin{figure}
		\includegraphics[width=0.45\textwidth]{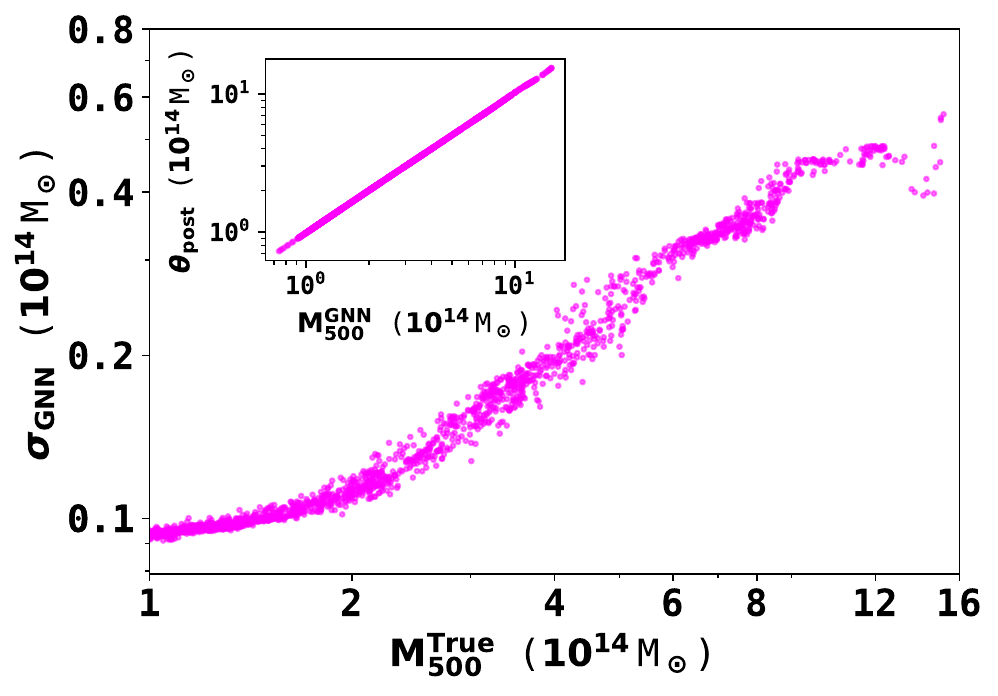}
                \centering

		\caption{\footnotesize The 1\(\sigma\) GNN model uncertainty, \( \sigma_{\mathrm{GNN}} \), estimated using KDE based SBI, is shown as a function of the true cluster mass \( {\rm M}_{500}^{\mathrm{True}} \). The inset figure shows that posterior means match the GNN output (point estimates), indicating a well-calibrated model.}
		\label{fig8b}
\end{figure}

\subsection{Model variance estimation}
To quantify model (epistemic) uncertainty in the inference of cluster masses, we adopt a simulation-based inference (SBI) framework \cite{2021MNRAS.501.4080K}. Following the approach of \cite{2022NatAs...6.1325D}, we employ a kernel density estimation (KDE) based SBI method in which the posterior distribution over the true cluster mass \( \boldsymbol{\theta} \) (i.e ${{\rm M}_{500}^{\rm True}}$) is inferred from the predicted mass \( \hat{\boldsymbol{\theta}} \) (i.e ${{\rm M}_{500}^{\rm GNN}}$) via a trained ensemble of GNNs. From the training set, we collect pairs of predicted and true masses \( (\hat{\boldsymbol{\theta}}_{\mathrm{train}}, \boldsymbol{\theta}_{\mathrm{train}}) \), and use them to estimate the joint distribution \( \mathcal{P}(\hat{\boldsymbol{\theta}}, \boldsymbol{\theta}) \) via a two-dimensional Gaussian KDE.

For a test sample with GNN predicted mass \( \hat{\boldsymbol{\theta}}_{\mathrm{test}} \), we obtain the conditional posterior \( \mathcal{P}(\boldsymbol{\theta} \mid \hat{\boldsymbol{\theta}}_{\mathrm{test}}) \) by evaluating and normalising the joint distribution at \( \hat{\boldsymbol{\theta}} = \hat{\boldsymbol{\theta}}_{\mathrm{test}} \). The posterior mean ($\boldsymbol{\theta}_{\mathrm{post}}$) and variance ($\sigma^2$) are then given by
\begin{align}
    \boldsymbol{\theta}_{\mathrm{post}} &= \int \boldsymbol{\theta} \, \mathcal{P}(\boldsymbol{\theta} \mid \hat{\boldsymbol{\theta}}_{\mathrm{test}}) \, d\boldsymbol{\theta} \nonumber \\
    \sigma^2(\boldsymbol{\theta} \mid \hat{\boldsymbol{\theta}}_{\mathrm{test}}) &= \int \|\boldsymbol{\theta} - \boldsymbol{\theta}_{\mathrm{post}}\|^2 \, \mathcal{P}(\boldsymbol{\theta} \mid \hat{\boldsymbol{\theta}}_{\mathrm{test}}) \, d\boldsymbol{\theta}.
\end{align}
This non-parametric approach avoids explicit likelihood modeling and enables flexible posterior estimation over cluster masses, grounded in the relationship between GNN predictions and true target values.

Figure \ref{fig8b} shows the predicted model uncertainty \( \sigma_{\mathrm{GNN}} \) versus the true cluster mass \( {\rm M}_{500}^{\mathrm{True}} \) for the test sample. We observe a clear trend of increasing uncertainty with increasing mass, suggesting that the model is less confident in its predictions for higher-mass clusters. This may be attributed to limited training examples at the high-mass end, or to more complex feature--target relationships in massive systems. However, the fractional percentage errors remain approximately $5\%$ for most clusters, increasing to about $10\%$ for clusters with masses below $2\times 10^{14} \, {\rm M}_{\odot}$. The inset plot shows posterior means closely match the GNN point estimates, indicating that the model is well-calibrated and that its predictions lie in regions of the input space well-supported by the training set. In later Sections, where observed data with errors are considered, we also incorporate aleatoric (data) uncertainty by performing Monte Carlo simulations to propagate these uncertainties through the model.

Finally, we note that probabilistic neural networks (e.g., evidential/Bayesian neural networks) were not employed in this work. First, these models often require strong parametric assumptions regarding the uncertainty structure and prior distributions. Second, our training sample is relatively small, particularly in the high-mass regime, which may hinder the precise estimation of model variance. Nonetheless, we plan to implement a full Bayesian formalism in future work with larger training datasets.
\begin{figure}
		\includegraphics[width=0.45\textwidth]{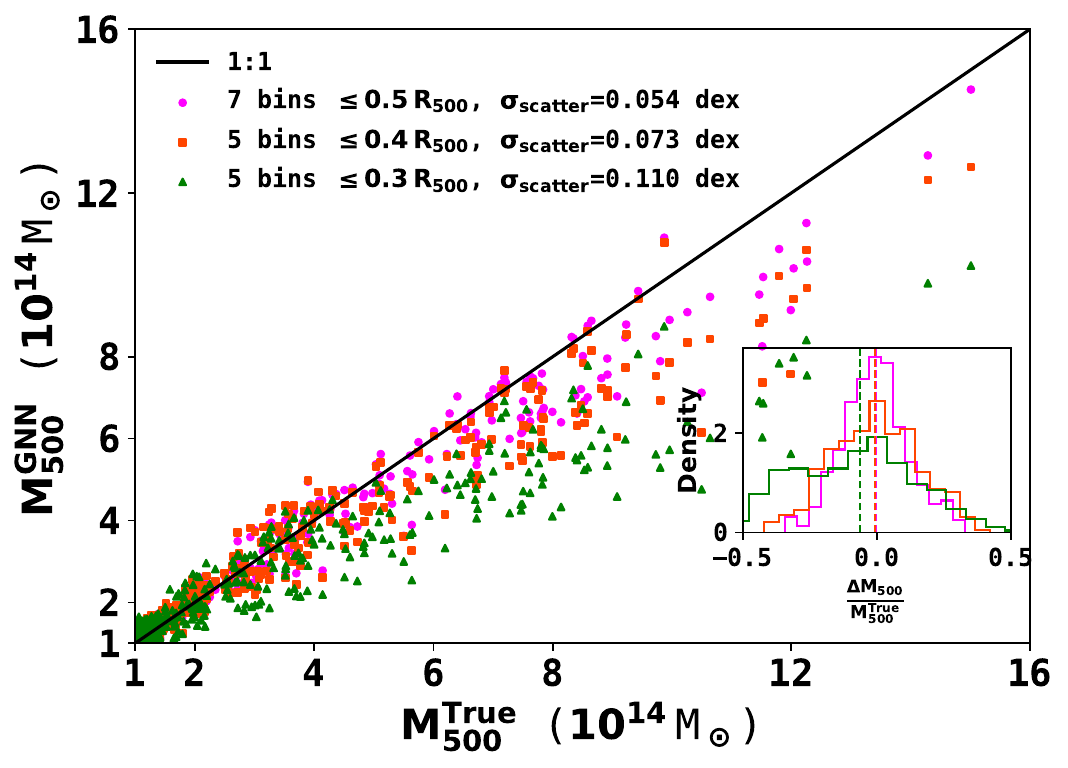}
                \centering

		\caption{\footnotesize Extrapolation performance of the GNN model when evaluated with input data restricted to the inner regions.
        The plot shows predicted versus true cluster masses for three cases where the available ICM data extends only up to 
        $0.5\,{\rm R}_{500}$, $0.4\,{\rm R}_{500}$, and $0.3\,{\rm R}_{500}$, corresponding to Case A, B, and C in the text and 7, 5, and 5 radial bins respectively.}
		\label{model_extra}
\end{figure}
\subsection{Model extrapolation ability}
\label{sub_A2}
To evaluate the extrapolation capabilities of our GNN model trained on ICM graph data spanning the radial range from $0.75\,{\rm R}_{500}$ to $2\,{\rm R}_{500}$, we conducted additional checks using ICM profiles restricted to smaller inner radial ranges using unagumented testing sample of 350 galaxy clusters. Specifically, we tested the model's performance when provided with data only up to $0.5\,{\rm R}_{500}$ (case A), $0.4\,{\rm R}_{500}$ (case B) and $0.3\,{\rm R}_{500}$ (case C) with 7, 5 and 5 bins respectively. Figure~\ref{model_extra} summarises the model predictions, while Table~\ref{tab:fitB} reports the best-fit linear relations between predicted and true masses for these cases. The median fractional residuals are $0.00 \pm 0.13$, $0.01 \pm 0.18$ and $0.06 \pm 0.25$ for cases A, B, and C, respectively. We observe an increasing scatter and a progressive bias in the recovered masses as the radial range decreases, indicating that the model’s performance degrades when extrapolating to the inner regions not covered during training.
\begin{figure}
		\includegraphics[width=0.45\textwidth]{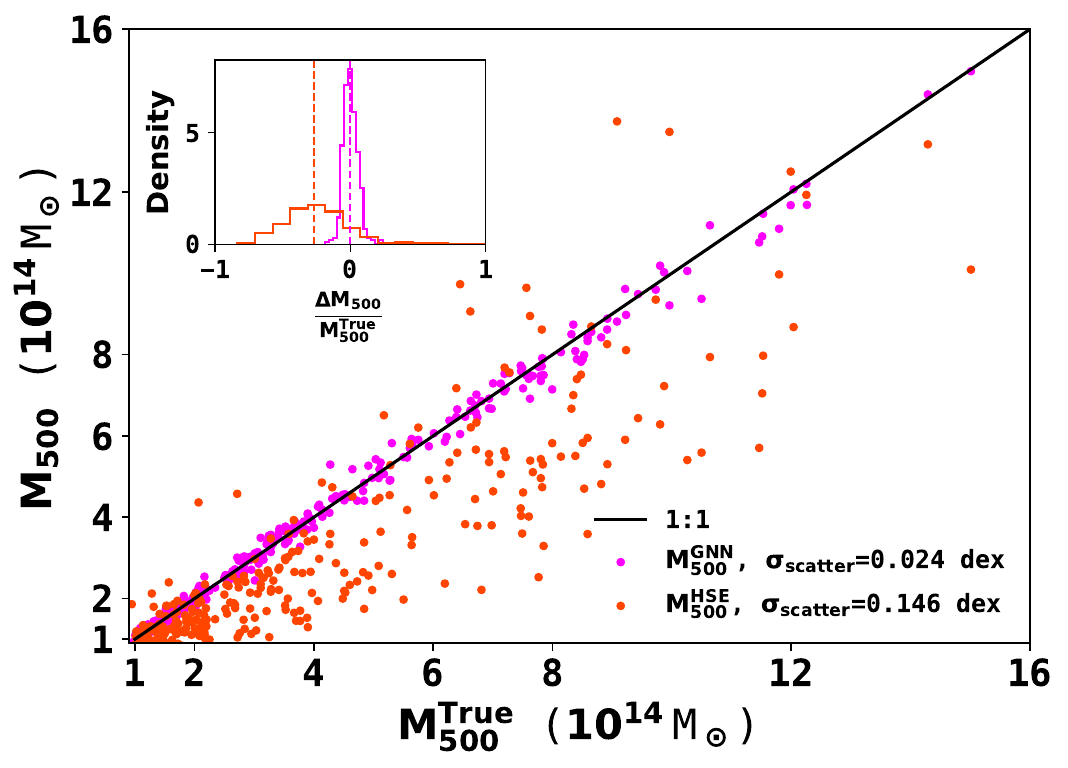}
                \centering

		\caption{\footnotesize Comparison of GNN and hydrostatic equation predictions for $\textrm{M}_{500}$ using an unaugmented test sample of 350 galaxy clusters. The black line represents a  1-to-1  relation. 
        Magenta and orange points respectively compare GNN predictions and hydrostatic predictions with true masses. The inset plot shows the distribution of residuals for both approaches, with the vertical 
        lines showing the median of the distributions.
		}
		\label{fig8}
\end{figure}
\begin{table}
    \centering
    \begin{threeparttable}
    \begin{tabular}{ccc}
        \toprule
        \textbf{Case} & \multicolumn{2}{c}{${\textrm M}^{\text{GNN}}_{500} = {\rm A} \times {\textrm M}^{\text{True}}_{500} + {\rm B} \times 10^{14} \, {\textrm M}_\odot$} \\
        \cmidrule(lr){2-3}
        & A & B \\
        \midrule
        7 bins up to $0.5\,{\rm R}_{500}$ & $1.119\pm0.011$ & $-0.230\pm0.047$ \\
       5 bins up to $0.4\,{\rm R}_{500}$ & $1.208\pm0.016$ & $-0.400\pm0.065$ \\
       5 bins up to $0.3\,{\rm R}_{500}$& $1.504\pm0.068$ & $-0.740\pm0.093$ \\
        \bottomrule
    \end{tabular}
    \caption{\footnotesize Best-fit relation between the GNN and true masses for different cases considered in Sec.~\ref{sub_A2}.}
    \label{tab:fitB}
    \end{threeparttable}
\end{table}
\subsection{Comparison to traditional hydrostatic masses}
The hydrostatic method, which assumes equilibrium within the ICM, typically relies on density and temperature (or pressure) measurements at a specific radius, R, to estimate the cluster mass within that radius. This approach, therefore, emphasises local indicators of mass, assuming a smooth, stable distribution of gas within the cluster. However, this assumption may not hold in clusters that have undergone significant dynamical interactions and in galaxy cluster groups, which may experience strong AGN-driven outflows to larger radii, leading to inaccuracies and increased scatter in mass estimates.

In contrast, the GNN model is more expansive, using both core and outer radial features in its mass predictions. The graph-based approach allows the GNN to capture the complex, non-local dependencies between different regions of the cluster, making it more robust to dynamical processes that can disrupt the ICM equilibrium. This capability allows the GNN to better capture the intricate dynamics of disturbed clusters, adapting to varying physical conditions. As shown in Fig.~\ref{fig8}, the scatter in $\textrm{M}_{500}$ estimates, in the unaugmented test sample, is larger for the hydrostatic method compared to the GNN model. We find that for the unaugmented test sample of 350 galaxy clusters, the median fractional dispersion in M$_{500}$ is biased by $-0.27\pm0.21$ and has a scatter that is about six times larger than that for the GNN-derived results. 
\begin{figure}
		\includegraphics[width=0.45\textwidth]{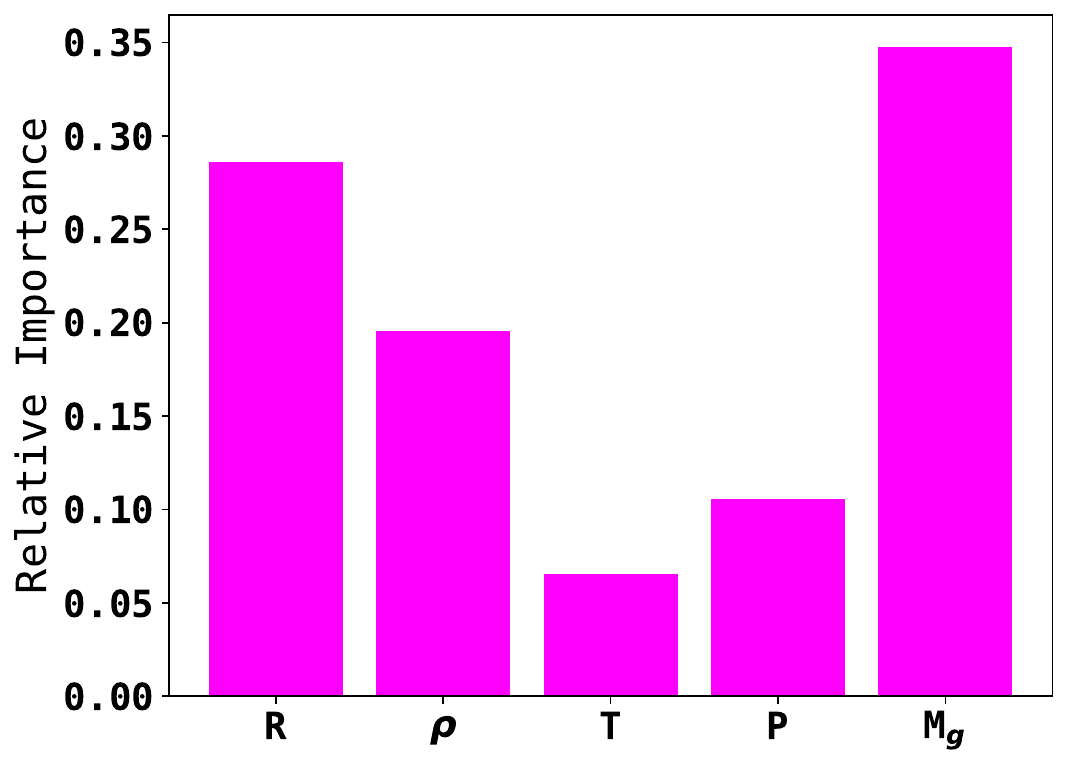}
                \centering
		\caption{\footnotesize Average feature importance for predicting M$_{500}$ as a function of input features. The relative importance of the five features i.e. density ($\rho$), temperature (T), pressure (P), gas mass (M$_g$), and radius (R).
		}
		\label{fig9}
\end{figure}
\subsection{Model interpretability: feature importance analysis}
We use integrated gradients to identify which features contribute most to the GNN mass predictions. This approach helps understand model behavior in complex astrophysical data, where feature contributions are not always clear. Integrated gradients calculate the gradients of the loss function with respect to each input feature by interpolating between a baseline (average feature profile) and the actual inputs. These gradients are tracked and averaged across batches, providing a measure of feature importance. Higher values indicate features that are more important for predicting galaxy cluster mass.

Figure~\ref{fig9} presents the feature importance for each input quantity in determining the total mass of galaxy clusters. The relative importance values are [R, $\rho$, T, P, M$_g$] = [0.29, 0.20, 0.06, 0.10, 0.34], highlighting that the model places the greatest emphasis on the integrated gas mass and radius features. This outcome is physically motivated: the gas mass profile, being a cumulative quantity, is a direct tracer of the gravitational potential. Moreover,  since the gas mass is derived from the integral of the gas density profile, it effectively encodes information regarding both the density normalisation and slope, which are key to determining the total mass distribution. The radius defines the spatial scale over which all other quantities (e.g., density, temperature) are evaluated, thus implicitly contributing to how gradients are computed. 

To test the relative importance of these features, we trained a model using only the radius and gas mass as inputs. Remarkably, the model performance showed a negligible decrease compared to when using the set of thermodynamic profiles used for the fiducial five feature model with dispersion and scatter of  $0.00\pm0.07$ and 0.030 dex, respectively. This result supports the physical interpretation with the integrated gradients analysis. However, its extrapolation capability is significantly compromised. When applied to clusters with limited radial coverage -- as in the extrapolation tests described previously -- the model performance, already poor in the fiducial case, deteriorates even further with scatter equal to 0.065, 0.090 and 0.130 dexs for the cases A, B and C, respectively. This suggests that the additional thermodynamic features, though not critical for performance within the training range, provide contextual information that is important for reliable generalisation beyond it.

In contrast, density, temperature, and pressure features - while central to understanding the thermodynamic state of the ICM -- do not provide as direct a constraint on the cluster mass if gas mass is taken into account. Their local values are sensitive to small scale fluctuations or non-thermal processes, making them less important predictors compared to gas mass. Therefore, the model naturally learns to prioritise integrative or structurally representative features -- M$_g$ and R -- which carry global information about the cluster mass distribution. However, as seen in Appendix~\ref{app0}, if the gas mass feature is ignored and the model is trained with the feature set consisting of radius, density, and temperature only, then density and radius become crucial, albeit with a larger scatter i.e. 0.038 dex for the testing sample.

\section{Application to observed data}
\begin{figure*}
    \centering
    \begin{minipage}{0.45\linewidth}
        \includegraphics[width=\textwidth]{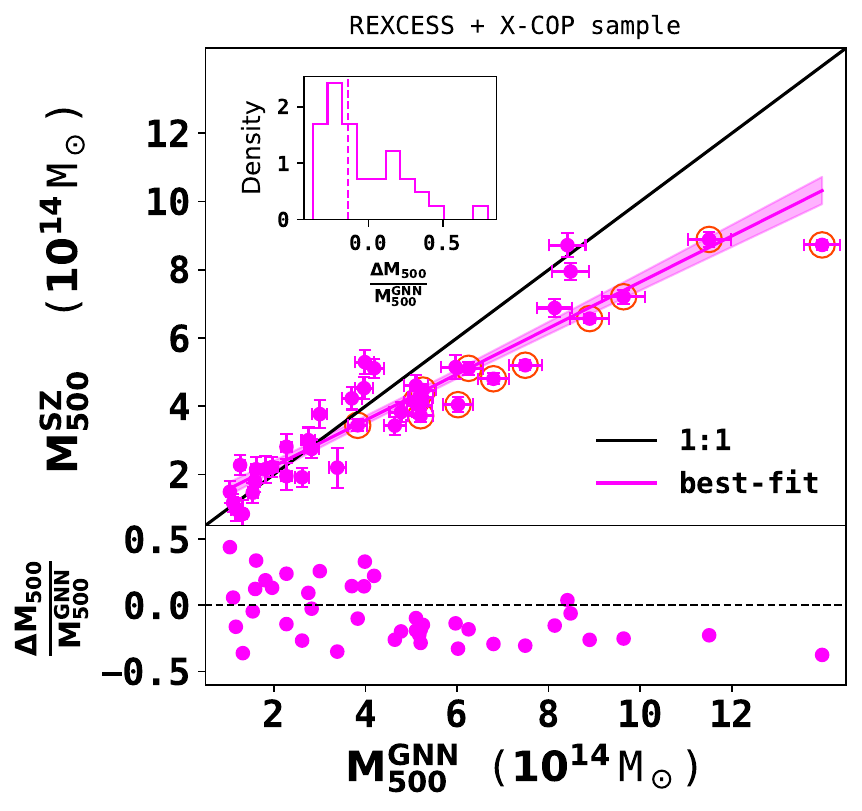} %
    \end{minipage}
    \begin{minipage}{0.45\linewidth}
        \includegraphics[width=\textwidth]{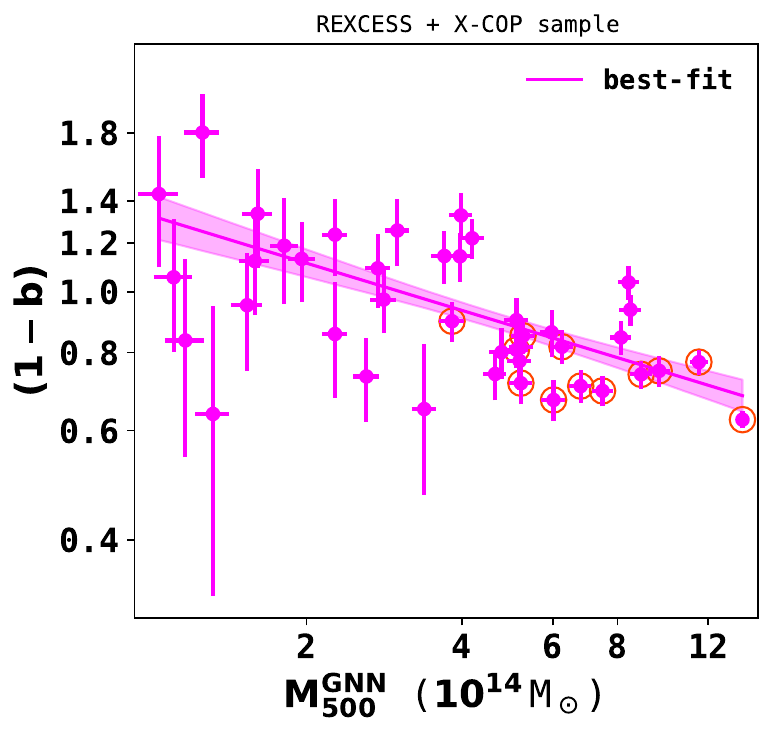}
    \end{minipage}
    \caption{\footnotesize Left panel shows the comparison of masses derived from the GNN model with those obtained from the SZ scaling relation. The black line represents the ideal 1-to-1 relation. The inset plots show the distribution of fractional residuals. The right panel shows the bias in the SZ-determined mass (M$^{\rm SZ}_{500}$/M$^{\rm GNN}_{500}$) as a function of GNN inferred mass. In both panels,  the magenta line and its associated shaded area indicate the best-fitting line and 1$\sigma$ region, respectively. The points enclosed with the orange circles belong to the X-COP sample. }
    \label{fig10} 
\end{figure*}

In this Section, we apply the GNN model to galaxy cluster data obtained from XMM-Newton X-ray observations. The X-ray observations that we will use and describe later provide all the features required by the GNN model, including the deprojected (3D) profiles that we have tested in the previous Sections.
This comparison provides insights into the potential improvements from machine learning models like GNN, and the need for corrections in mass scaling relations.

The SZ cluster mass, M\(_{500}^{\textrm{SZ}} \), is derived from the SZ effect, which measures the interaction of high-energy electrons in the ICM with the CMB photons. This interaction distorts the CMB spectrum, with the SZ observable, \( Y_{\textrm{SZ}} \propto \int \textrm{P} dV \), representing the volume-integrated pressure of the ICM. The SZ effect, being independent of redshift, provides a direct and robust probe of cluster properties across cosmic epochs. The SZ scaling relation is established by calibrating these SZ and X-ray observables against hydrostatic mass estimates, M\(_{\textrm{HSE}} \), derived from detailed modeling of pressure and density profiles \citep{2010A&A...517A..92A}. Such scaling relations are indispensable for cluster-based cosmological studies, enabling robust constraints on dark energy and structure formation. For cluster cosmology using the Planck SZ derived masses as reference cluster masses \citep{2013A&A...550A.129P}, it has been demonstrated that the SZ and X-ray derived masses are generally consistent within expected uncertainties \citep{2013A&A...550A.131P}\footnote{We note that these observed M$_{500}$ are computed with $H_0=70$ km s$^{-1}$ Mpc$^{-1}$. Consequently, for the same profiles, the masses derived from the GNN model, which assumes $H_0=67.7$ km s$^{-1}$ Mpc$^{-1}$ cosmology,  are expected about 1\% larger than M\(_{500}^{\textrm{SZ}} \) and M\(_{500}^{\textrm{X}} \) values, assuming no bias. }.  In Appendix~\ref{app2} and Appendix~\ref{app3}, we also compare the X-ray derived masses with the GNN masses.

For our analysis, we consider two well-known cluster samples. We first consider the REXCESS (Representative XMM-Newton Cluster Structure Survey) sample, which consists of a carefully selected set of 31 galaxy clusters from the REFLEX catalogue, having redshifts between $z$ = 0.055 and 0.183. These clusters were chosen to provide a statistically representative view of cluster properties across a range of masses and morphologies. Using high-quality X-ray observations from XMM-Newton has resulted in in-depth analysis of the ICM in these clusters  
 \citep{Bohringer2007,cro08,pra07,Pratt2010, 2010A&A...517A..92A}. This analysis includes deriving radial profiles of density, temperature, and entropy, which are crucial for understanding the thermodynamic state and structural characteristics of the clusters. Studying these profiles and establishing scaling relations, such as those between X-ray luminosity and temperature, has led to valuable insights into the physics governing galaxy clusters and aids in cosmological studies that use clusters as probes of large-scale structure and cosmic evolution. In our previous work, we used this sample to study the non-gravitational feedback in galaxy clusters \citep{Iqbal2018}.  
 
 We also consider the X-COP (XMM-Newton Cluster Outskirts Project) sample which is an X-ray (XMM-Newton) follow-up project of a {\it Planck} SZ-selected set of 12 galaxy clusters in the redshift range of  0.04 $< z <$ 0.1, chosen to facilitate a detailed study of cluster properties extending into their outer regions \citep{2017AN....338..293E}. The selection criteria leverage the SZ effect’s sensitivity to gas pressure and the X-ray emission’s dependence on gas density, enabling comprehensive measurements of density, pressure, temperature, and entropy profiles from the cluster core to the outskirts \citep{2019A&A...621A..41G,2019A&A...621A..40E}.  The X-COP sample's findings contribute to a better understanding of galaxy cluster formation, as well as refining cluster-based cosmological constraints by reducing biases associated with mass estimates \citep{2019A&A...621A..40E,2022A&A...662A.123E}. In particular, \cite{2019A&A...621A..40E} not only estimated the hydrostatic mass but also proposed a method to estimate the total mass, which includes the contribution from non-thermal pressure using hydrodynamical simulations \citep{2014ApJ...792...25N}.
 
 \subsection{Estimation of GNN masses (M$_{500}^{\rm GNN}$)}
 We use the ICM profiles of the REXCESS sample derived in \cite{Bohringer2007,cro08,Pratt2010,2010A&A...517A..92A} to estimate M$_{500}$ using the GNN model. Similarly, for the X-COP sample, thermal profiles derived in \cite{2019A&A...621A..41G} were used to estimate GNN masses. As the GNN model output is predominantly influenced by the gas mass, we employ density-defined radial annuli at the node points and interpolate the temperature profile to estimate its values at these nodes. However, only the nodes whose radii fall within the observed range of the temperature profile are taken into account. Moreover, the innermost 50 kpc of the profiles were excluded, since this region is dominated by feedback processes. To obtain an estimate of the uncertainty in the GNN masses, we simulated  100 observed ICM profiles by assuming  log-normal distribution for density and temperature profiles \citep{2007ApJ...659..257K}, with the standard deviation taken as the observed uncertainties. For each cluster, we computed GNN derived mass,  M\(_{500}^{\textrm{GNN}}\), as the mean output of these 100 simulations. Two contributions to the total uncertainty were then considered. The first is the variance of the predicted masses across the 100 realisations, which reflects the sensitivity of the GNN predictions to observational uncertainties in the input profiles. The second is the mean of the SBI predicted variances from the GNN model for each realisation, which represents the intrinsic uncertainty of the model predictions. Tables \ref{appx1} and \ref{appx2} in appendices \ref{app2} and \ref{app3} provides the GNN mass for the REXCESS and X-COP samples, respectively, along with other mass proxies. We emphasise that the uncertainties in the GNN derived cluster masses are primarily driven by model variance, rather than by the observational uncertainties of the input X-ray profiles.
 
 \subsection{Estimation of SZ masses (M$_{500}^{\rm SZ}$)}\label{sec:szmass}
Out of the 31 clusters in the REXCESS sample, 23 are included in the PSZ2 cluster catalogue~\citep{PlanckClusterCatalogue2016}, meaning the rest are below the PSZ2 catalogue detection threshold corresponding to a signal-to-noise of 4.5.
We thus lacked published SZ mass estimates for a large fraction of the sample. To overcome this problem, we performed a targeted, unblind extraction of the REXCESS sample in the {\it Planck} data, fixing the position to that of the REXCESS catalogue. We also performed the same extraction for the twelve X-COP clusters to obtain fully homogeneous SZ mass estimates of the REXCESS+X-COP sample.

In practice, we worked with the six all-sky maps of the High Frequency Instrument of {\it Planck}~\citep{HFI2016}. We used the MMF3 algorithm based on multi-frequency matched filters developed for the {\it Planck}  analysis~\citep{Melin2006,Melin2012}.
The MMF3 algorithm produces cutouts ($10 \times 10 \, {\rm deg}^2$) of the {\it Planck} maps centered on each cluster and combines them optimally to extract the cluster flux. The algorithm assumes a fixed spatial profile for the SZ signal, specifically, the universal pressure profile of~\citet{2010A&A...517A..92A} while leaving the amplitude as a free parameter. The SZ spectral dependence was also modeled using the non-relativistic limit. The MMF3 algorithm scans over a grid of 32 logarithmically-spaced $\theta_{500}$\footnote{Projected cluster radius enclosing the mass M$_{500}$.} from $0.94$ to $35.32 \, {\rm arcmin}$. For each cluster, we extracted its SZ flux $Y_{500}$ as a function of the size $\theta_{500}$ and the associated signal-to-noise ratio $S/N$. Following the method described in Sect.~7.2.2 of~\cite{PlanckClusterCatalogue2014}, we broke the  Planck flux--size degeneracy by applying an X-ray prior based on the scaling relation given in Eq.~(5) of~\cite{PlanckClusterCatalogue2014}, which links the SZ flux to the cluster size, assuming the redshift is known. This allowed us to derive the {\it Planck} flux and, subsequently, the SZ mass M\(_{500}^{\rm SZ} \) of the cluster. The corresponding 68\% confidence intervals on the flux were estimated using the same approach, by calculating \( Y_{500} \pm Y_{500}/(S/N) \).
We were able to extract homogeneous flux and masses for all the clusters but one, RXC J1236.7--3354, from REXCESS, whose signal to noise was $S/N=0.2$, resulting in a measured flux consistent with zero.


\subsection{Comparison of SZ masses with GNN masses}
To analyse the comparison between masses derived from the GNN model and those obtained from the SZ scaling relation, we can focus on the bias and the fractional residuals as key metrics. 
Figure~\ref{fig10} shows the comparison of GNN masses to the SZ masses and the relative bias (M$^{\rm SZ}_{500}$/M$^{\rm GNN}_{500}$) for the combined REXCESS and X-COP sample. Assuming M\(_{500}^{\textrm{GNN}}\) to be the `true' unbiased cluster masses, we find a bias $(1-b)$ and fractional residuals of $0.81^{+0.34}_{-0.14}$  and $-0.14^{+0.27}_{-0.20}$ respectively for M$^{\textrm{SZ}}_{500}$ masses. However, if we only consider the X-COP sample we find the bias and fractional residuals to be $0.74^{+0.08}_{-0.06}$ and  $-0.25^{+0.08}_{-0.06}$ respectively. The significantly tighter constraints observed in the X-COP sample could be attributed to its better data quality and to the sample's extended coverage into the cluster outskirts. In contrast, the REXCESS sample comprises relatively lower-mass clusters, as seen in Fig.~\ref{fig10}, and tends to exhibit smaller mass biases and in fact a bias of >1 is observed for many low mass clusters. This contributes to the broader dispersion observed in the combined sample.  We find the best-fitting linear relation for the combined sample by employing the Gaussian mixture model {\tt LinMix}  \citep{2007ApJ...665.1489K} 
\begin{eqnarray}
{\textrm M}^{{\text{SZ}}}_{500} = (0.67 \pm 0.04) \, {\textrm M}^{\text{GNN}}_{500} +(0.89\pm 0.22) \times 10^{14}{\textrm M}_\odot.
\end{eqnarray}
In particular, if the intercept is fixed to zero, we find a best-fitting slope (bias) to be $0.81\pm0.02$.

The right panel of Fig.~\ref{fig10} shows the relation between the bias and cluster mass. We quantify the bias-mass relationship by the following power-law fit
\begin{equation}
(1-b)=(0.84\pm0.03) \times \left(\frac{\textrm{M}_{500}^{\textrm{GNN}}}{6\times10^{14}{\textrm M}_\odot}\right)^{-0.25\pm 0.05}.
\end{equation}
This relation reveals a strong trend, at 5$\sigma$ significance, of decreasing $1-b$ (i.e. increasing bias) with increasing cluster mass, indicating that more massive clusters exhibit stronger biases in SZ derived masses. This trend is physically consistent with the expectation that non-thermal pressure support arising from bulk motions, turbulence, and ongoing merger activity becomes increasingly important in massive systems. 

\cite{2022NatAs...6.1325D} also found that less massive clusters exhibit a smaller bias when comparing {\it Planck} masses to those estimated with a deep learning technique using the \thethree\, simulation. They further argued that the derived bias could be potentially explained due to differences in the slopes of $Y$-M scaling relations between the observations and \thethree \, simulations. This explanation aligns with the positive bias found in some of the low mass clusters in our work.

A comprehensive assessment that incorporates additional non-SZ mass proxies -- such as weak lensing, galaxy dynamics, or refined hydrostatic mass estimates which incorporate no-thermal pressure -- would provide a more robust cross-validation of our results. Each of these methods probes different physical aspects of the cluster potential: weak lensing directly traces the total matter distribution without assumptions about the dynamical state, galaxy dynamics capture the kinematic imprint of the gravitational potential, and hydrostatic masses constrain the thermodynamic equilibrium of the ICM. By jointly considering these complementary techniques, one could better quantify systematic uncertainties (e.g., non-thermal pressure support, line-of-sight projection effects, or sample selection biases) and isolate potential deviations from the self-similar scaling expectations. However, such an analysis requires careful homogenisation of datasets and treatment of selection effects, which is beyond the scope of the present work. We therefore defer this more detailed, multi-probe study to future work.

\section{Discussion and Conclusions}
Hydrodynamical simulations of galaxy clusters are essential for interpreting X-ray observations, as they model the complex distribution and behavior of the hot and diffuse ICM. These simulations enable the prediction and comparison of projected quantities such as temperature, density, pressure, and entropy profiles, which are critical for analysing the physical conditions within clusters. By matching these projections to observed X-ray data, we can gain insights into key processes, such as gas cooling, shock heating, turbulence, and feedback from AGN. Recent advances in hydrodynamical simulations have led to increasingly realistic models that closely match observations in both large-scale structures and finer intra-cluster features \citep{cui2018,2024MNRAS.533.2656B,2024A&A...687A.129L,2024A&A...691A.340R}.

On the other hand, deep learning techniques have become increasingly important for unveiling complex non-linearities in astrophysical data.  While traditional methods may struggle to capture the intricate relationships between different physical quantities, especially when dealing with large, noisy observations or when attempting to model highly non-linear processes, deep learning models have shown great promise in learning these complex relationships directly from observational data without relying heavily on pre-defined physical assumptions \citep{Alzubaidi2021ReviewOD}.

Departing from traditional methods that rely on the hydrostatic equilibrium assumption, our approach incorporates high-resolution thermal profiles of the ICM and total dark matter profiles from hydrodynamical simulations and a deep-learning approach to uncover the intricate nature of the ICM and cluster mass (M$_{500}$) relationship using X-ray observations, implicitly accommodating the effects of non-thermal pressure components. While the estimates of integrated mass, M(<R), at a given radius, R, in the hydrostatic framework rely heavily on data around R and are sensitive to noisy derivatives, our approach learns the overall structure of the data to yield robust estimates of M$_{500}$. We find that, in general, our GNN model can accurately recover M$_{500}$. We summarise our results and the performance of the  GNN model in estimating the mass of galaxy clusters as follows:
\begin{itemize}
\item The performance of the trained GNN model is consistent across both the training and testing sets, with a bias in $M_{500}$ equal to zero and a median fractional dispersion of around 6\% within the mass range $1 \times 10^{14}\, \textrm{M}_\odot \leq \textrm{M}_{500} \leq 2 \times 10^{15}\, \textrm{M}_\odot$.  Additionally, our model outperforms classical hydrostatic modeling, which shows a bias of approximately 20\% as well as scatter that is about six times larger than that of our GNN based approach.

\item We evaluate the GNN performance on the most relaxed and disturbed clusters in the test sample to understand how cluster dynamical state affects model adaptability. The GNN achieves consistent adaptability across these morphological states, with comparable error distributions between relaxed and disturbed clusters. This suggests the model effectively captures key structural features of the ICM, maintaining stable performance despite variations in cluster dynamical states.

\item We tested the model using observational-like binning, truncating the profiles at R$_{500}$ and $0.75\,{\rm R}_{500}$ (15 and 7 bins). Despite the reduced radial coverage, the GNN estimates cluster masses with high accuracy, yielding scatter of 5.6\%, 5.9\% and 9.1\% for the 48, 15 and 7 bin cases, respectively. Moreover, if the core input profiles are excluded (<0.1R$_{500}$), the scatter in the 7 bin case is  reduced to 6.4\%

\item We use integrated gradients to reveal which features most influence the GNN's mass predictions. We find that ICM gas mass and radius are the most influential features, as they directly correlate (increasing profiles) with the galaxy cluster's mass. Other features, such as temperature and pressure, play supporting roles but have a less direct impact on mass prediction.

\item We applied the GNN model to the joint REXCESS and X-COP  X-ray sample. Our finding indicate a strong mass dependent bias, at $5\sigma$ confidence, with high mass clusters exhibiting a greater degree of bias.

\item The significant detection of a mass-dependent bias could introduce a crucial refinement to the use of galaxy clusters as cosmological probes. Such a trend implies that high mass clusters typically used as anchors in cosmological analyses may be more strongly biased, potentially leading to the reduction in the tension in \( \sigma_8 \) between the CMB cosmology and SZ cluster counts.
\end{itemize}

Our results align with recent neural network models developed in \cite{2022NatAs...6.1325D,2022arXiv220712337F,2024MNRAS.528.1517D}, which successfully learn the connections between observed maps and total mass maps or profiles, providing unbiased estimates of total mass. However, the GNN approach introduced in this work significantly diverges from these methodologies. While the studies mentioned rely on fixed-grid, image-based deep learning techniques such as CNNs and U-Net to estimate mass proxies or project total mass density maps from multi-wavelength hydrodynamic simulations having fixed-grid  (with \citealt{2022NatAs...6.1325D} exclusively using SZ maps to estimate M$_{500}$), the GNN model processes 3D deprojected radial profiles of the intracluster medium obtained in X-rays. This is achieved by learning the spatial and hierarchical structure of data encoded as a graph. Although the GNN approach assumes a simplified spherical symmetry and uses 3D profiles that are not directly observable, it achieves better accuracy while offering the flexibility to handle input data of varying sizes.

Our models used 3D profiles from simulations and are tested on deprojected profiles from observations. Potentially, one could use 2D features, that is, the projected observational profiles. However, this would require training separate models that we aim to explore in the future, even extending the work to different instruments, thus changing the response matrices and point-spread functions.

The choice of simulation suite can have an impact on the model's ability (in fact, this is likely the main source of uncertainty in our results) to accurately predict cluster masses, as different simulation codes may incorporate varying physical models, resolution, and assumptions about dark matter, baryons, and feedback mechanisms \citep{2021ApJ...915...71V,2023MNRAS.522.2628W}. 
Cross-simulation testing -- where a model trained on one simulation (e.g., GADGET-X) is validated on another (e.g, GIZMO-SIMBA \citep{2022MNRAS.514..977C} -- can help identify systematic biases introduced by simulation-specific physical prescriptions.  Additionally, integrating weak gravitational lensing mass estimates as a complementary observational tool can be invaluable in mitigating simulation-related biases. This type of comparison can also guide improvements in both the simulation models and the machine learning models, ensuring more accurate mass predictions for cosmological studies and cluster-based cosmology.

Our future goal is to develop a more robust GNN model by incorporating a larger simulated dataset and integrating multi-wavelength observations. While a comprehensive comparison across multiple simulation sets lies beyond the scope of the present study, we acknowledge the importance of this factor and plan to address it in future work by applying our method to alternative simulation suites with different physical prescriptions. Furthermore, \cite{2024MNRAS.528.1517D} demonstrated that including stellar mass maps significantly improves the estimates of projected total mass profiles in galaxy clusters. We aim to enhance our model by incorporating Compton-$y$ and stellar mass profiles as well as test the impact of other features such as X-ray luminosity and photon count profiles. Another important extension of our project will be to explore the capability of the GNN model to estimate the radial total mass profiles in galaxy clusters, instead of estimating just M$_{500}$.

Advanced techniques such as the deep learning approach proposed in this paper are useful to account for and mitigate the impact of the biases arising from the hydrostatic equilibrium assumption and the scaling relations derived therefrom. In the future, we expect such approaches to offer a powerful tool for accurately determining galaxy cluster masses, and hence contributing towards a more robust cluster cosmology. 
\section*{Data Availability}
The GNN model developed in this work is available on request from the corresponding author. The simulations used to train the model are part of \thethree\footnote{\url{https://www.nottingham.ac.uk/astronomy/The300/index.php}}. Results from the REXCESS project, including detailed X-ray observation profiles, are published in \citet{Bohringer2007,cro08, Pratt2010, 2010A&A...517A..92A} and were kindly provided by the project team. The X-COP results, which focus on combined X-ray and SZ observations of massive galaxy clusters, are publicly accessible online\footnote{\url{https://dominiqueeckert.wixsite.com/xcop/about-x-cop}}.

\begin{acknowledgements}
 The simulations were performed at the MareNostrum Supercomputer of the BSC-CNS through The Red Espa\~nola de Su\-percomputaci\'on grants (AECT-2022-3- 0027, AECT-2023-1-0013), and at the DIaL – DiRAC machines at the University of Leicester through the RAC15 grant: Seedcorn/ACTP317. DA thanks the Ministerio de Ciencia e Innovación (Spain) for financial support under Project grant PID2021-122603NB-C21 and Atracci\'{o}n de Talento Contract no. 2020-T1/TIC-19882 granted by the Comunidad de Madrid in Spain. WC gratefully thanks Comunidad de Madrid for the Atracci\'{o}n de Talento fellowship No. 2020-T1/TIC19882 and Agencia Estatal de Investigación (AEI) for the Consolidación Investigadora Grant CNS2024-154838. He further acknowledges the Ministerio de Ciencia e Innovación (Spain) for financial support under Project grant PID2021-122603NB-C21, ERC: HORIZON-TMA-MSCA-SE for supporting the LACEGAL-III (Latin American Chinese European Galaxy Formation Network) project with grant number 101086388 and the support by the China Manned Space Program with grant no. CMS-CSST-2025-A04. SM and AI acknowledge support of the Department of Atomic Energy, Government of India, under project no. 12-R\&D-TFR-5.02-0200. GWP acknowledges long-term supports from CNES, the French space agency. We gratefully acknowledge the support of the GPU-equipped High-Performance Computing resources at the University of Lille for enabling the computational aspects of this research. The authors thank the anonymous referee for their careful reading of the manuscript and insightful comments, which significantly improved the clarity and quality of this work.
 \end{acknowledgements}

\bibliographystyle{aa}
\bibliography{example} 

\appendix
\section{Simulated ICM profiles in \thethree}
\label{app00}
\begin{figure*}
		\includegraphics[width=0.90\textwidth]{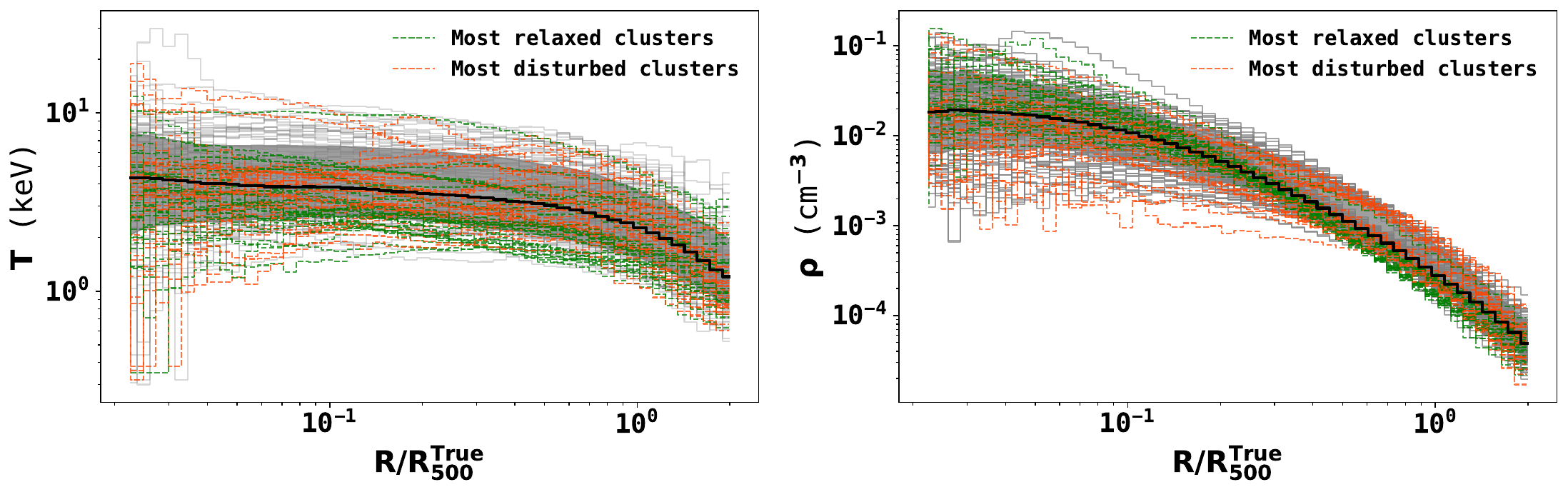}
        \centering
		\caption{\footnotesize ICM temperature (left) and density (right) profiles of 100 randomly-selected simulated galaxy clusters in \thethree. The grey lines show individual profiles and the green and orange lines show the 30 most relaxed and disturbed clusters, respectively, according to criteria given by $\chi$ (see text for details).  The solid black line shows the median of the full simulated sample.
		}
		\label{fig1}
\end{figure*}

Figure~\ref{fig1} presents the ICM temperature and density profiles for a subset of 100 randomly-selected simulated galaxy clusters in \thethree\ dataset.
\section{Summary of GNN model architecture and training Hyper-parameters}
\label{app000}
Here we provide additional details on the GNN architecture and the hyper-parameters used for training.
Table~\ref{tab:gcn_architecture} outlines the full architecture of the GNN model, including the type of each layer, the dimensionality of its input and output, and a brief description of its role in the pipeline.
Table~\ref{tab:hyperparameters} lists the optimised hyper-parameters used during training, including the learning rate, regularisation settings, and training schedule.

    \begin{table*}
    \scriptsize

    \centering
    \begin{threeparttable}
    \begin{tabular}{ccl}
      \toprule
\toprule
    
        \textbf{Hyper-parameter} &\textbf{Range/Values considered}& \textbf{Best-fit}  \\
        \midrule
        Learning rate & $10^{-6}-10^{-2}$& \(0.00099\) . \\ \hline
        Weight decay & $10^{-3}-10^{-1}$&  \(0.0029\)  \\ \hline
        Learning rate factor& [0.1, 0.3, 0.5, 0.7, 0.9] & \(0.5\)  \\ \hline
        Embedding size &[64, 100, 128]& \(128\)  \\ \hline
        Dropout probability ($p$) &0 & Fixed \\ \hline
        Batch size & [32, 64, 128]& \(128\)  \\ \hline
        Optimiser & {\tt Adam} & Fixed. \\ \hline
        Scheduler & {\tt ReduceLROnPlateau} & Fixed \\ \hline
      GNN layers& 1-4 & 3  \\ \hline
      GNN layer type&   {\tt[GCNConv, GATConv, SAGEConv, TransformerConv} ] & {\tt TransformerConv} \\ \hline
      Pooling Type &[ {\tt Mean, Add, Max, Attention}] &  {\tt Max}\\ \hline
      Linear Layers & 3 &  Fixed\\ \hline

        Epochs & \(500\) & Fixed \\
       \bottomrule
    \end{tabular}
    \caption{\footnotesize Optimised hyper-parameters for our fiducial 5 feature GNN model.}
    \label{tab:hyperparameters}
    \end{threeparttable}
\end{table*}

\begin{table*}
\scriptsize
    \centering
    \begin{threeparttable}
    \begin{tabular}{c c  c c p{9cm}}   \toprule
\toprule
        \textbf{Layer} & \textbf{Type} & \textbf{Input size } & \textbf{Output size} & \textbf{Description} \\

         & & \textbf{Nodes $\times$ Features} & \textbf{Nodes $\times$ Features} &\\
       \midrule 
        1 & TransformerConv & $n \times 5$ & $n \times 128$ & First GNN layer transforming input with $n$ nodes of 5-dimensional features to 100-dimensional embedding for each node. $n$ is between 7-48 \\
        \hline
        2 & BatchNorm1d & $n \times 128$ & $n \times 128$ & Normalises the output of the first GNN layer. \\
        \hline
        3 & ReLU & $n \times 128$ & $n \times 128$ & Applies ReLU activation. \\
        \hline
        4 & TransformerConv & $n \times 128$ & $n \times 128$ & Second GNN layer refining 100-dimensional embeddings for each node. \\
       \hline
        5 & BatchNorm1d & $n \times 128$ & $n \times 128$ & Normalises the output of the second GNN layer. \\
        \hline
        6 & ReLU & $n \times 128$ & $n \times 128$ & Applies ReLU activation. \\
        \hline
        7 & TransformerConv & $n \times 128$ & $n \times 128$ & Third GNN layer refining 100-dimensional embeddings for each node. \\
       \hline
        8 & BatchNorm1d & $n \times 128$ & $n \times 128$ & Normalises the output of the second GNN layer. \\
        \hline
        9 & ReLU & $n \times 128$ & $n \times 128$ & Applies ReLU activation. \\
        \hline
        10 & Global pooling & $n \times 128$ & $1 \times 128$ & Aggregates node features into a single vector via global max pooling. \\
        \hline
        11 & Concatenation & $(1 \times 128) + (1 \times 1)$ & $1 \times 129$ & Combines GAP output with an external feature (e.g., redshift). \\
        \hline
        12 & Linear & $1 \times 129$ & $1 \times 60$ & Maps concatenated features to 60 dimensions. \\
        \hline
        13 & BatchNorm1d & $1 \times 60$ & $1 \times 60$ & Normalises the output of the first linear layer. \\
        \hline
        14 & ReLU & $1 \times 60$ & $1 \times 60$ & Applies ReLU activation. \\
        \hline
        15 & Linear & $1 \times 60$ & $1 \times 30$ & Maps 60 features to 30 dimensions. \\
        \hline
        16 & BatchNorm1d & $1 \times 30$ & $1 \times 30$ & Normalises the output of the second linear layer. \\
        \hline
        17 & ReLU & $1 \times 30$ & $1 \times 30$ & Applies ReLU activation. \\
        \hline
        18 & Linear & $1 \times 30$ & $1 \times 1$ & Final layer that outputs a single prediction value. \\
      \bottomrule

    \end{tabular}
    \caption{\footnotesize Summary of the GNN architecture for our fiducial 5 feature model used in this work, with matrix dimensions for each layer.}
    \label{tab:gcn_architecture}
        \end{threeparttable}

\end{table*}

\section{Core-excised GNN modeling}
\label{sec:Core-excised GCN modeling}
Core-excised regions in galaxy clusters are defined by excluding the innermost areas, typically within $0.15\,\textrm{R}_{500}$, to minimise the impact of non-gravitational processes such as radiative cooling, AGN feedback, and mergers. \cite{2022A&A...665A..24P} showed that these central processes introduce significant variability in the gas density, leading to a large intrinsic scatter of approximately $40\%$. Moreover, beyond the core, they found that thermal profiles evolve more self-similarly under the influence of gravity, resulting in a reduced scatter of around $20\%$ at intermediate radii [$0.5$--$0.7]\,\textrm{R}_{500}$. For example, the core-excised X-ray luminosity ($L_{Xc}$), measured within the [$0.15$--$1]\,\textrm{R}_{500}$ annulus, exhibits a tight correlation with cluster mass, with a low logarithmic scatter of $\sim13\%$. Similarly, \cite{2023MNRAS.518.2735I} demonstrated that the AGN luminosity required to significantly influence the ICM out to $\sim[0.1-0.2]\,\textrm{R}_{500}$ is comparable to the observed radio jet power, reinforcing the view that AGN feedback is the dominant heating mechanism in cluster cores.

To investigate the effect of core exclusion on mass estimation using machine learning, we also trained a GNN model without the core regions. In the core-excised setting, input features were chosen randomly such that the first data point is within the radial range [0.1-0.2]$\,\textrm{R}_{500}$ and the last data point is within [0.75-2]$\,\textrm{R}_{500}$, effectively excluding the central ICM region. Our results suggest that removing the core profiles only leads to a slight reduction in scatter for the high resolution profiles (i.e 48 bins input radial profiles).  We find for the unaugmented testing sample (350 clusters), the median bias and fractional residuals to be $1.00\pm0.06$ and $0.00\pm0.05$ respectively and a scatter of 0.022 dex. However, for the observationally motivated 15 bin and 7 bin cases, for this sample, we find improved scatters of 0.023 dex and 0.024 dex, respectively, compared to the fiducial model, where the scatter for the 7 bin case was 0.036 dex.  Finally, we find for this case,  feature importance values to be  [R, $\rho$, T, P, M$_g$]=[0.28, 0.19, 0.06, 0.1, 0.37].

\section{Variants of GNN model having different feature sets}
\label{app0}
Here, we present two variants of the GNN models, differing in the number of input features. The lower dimensional model incorporates only the density and temperature of the ICM, along with the associated radius, providing a simpler representation of the system. In contrast, the higher dimensional model includes entropy as an additional feature, alongside the five other features considered in the fiducial model.\footnote{The entropy of the ICM is defined as $\rm K = k_bT / n_e^{2/3}$, where T is the temperature, k$_{\rm b}$ is the Boltzmann’s constant and n$_e$ is the electron density.}

\begin{figure*}
    \centering
    \begin{minipage}{0.45\linewidth}
        \includegraphics[width=\textwidth]{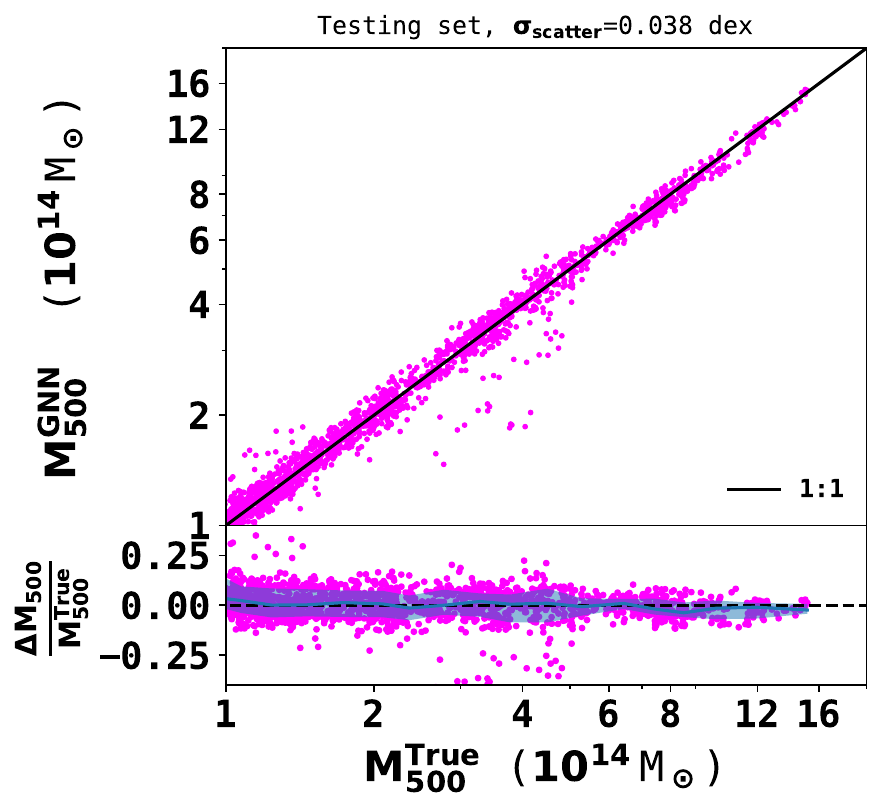} %
    \end{minipage}
    \begin{minipage}{0.45\linewidth}
        \includegraphics[width=\textwidth]{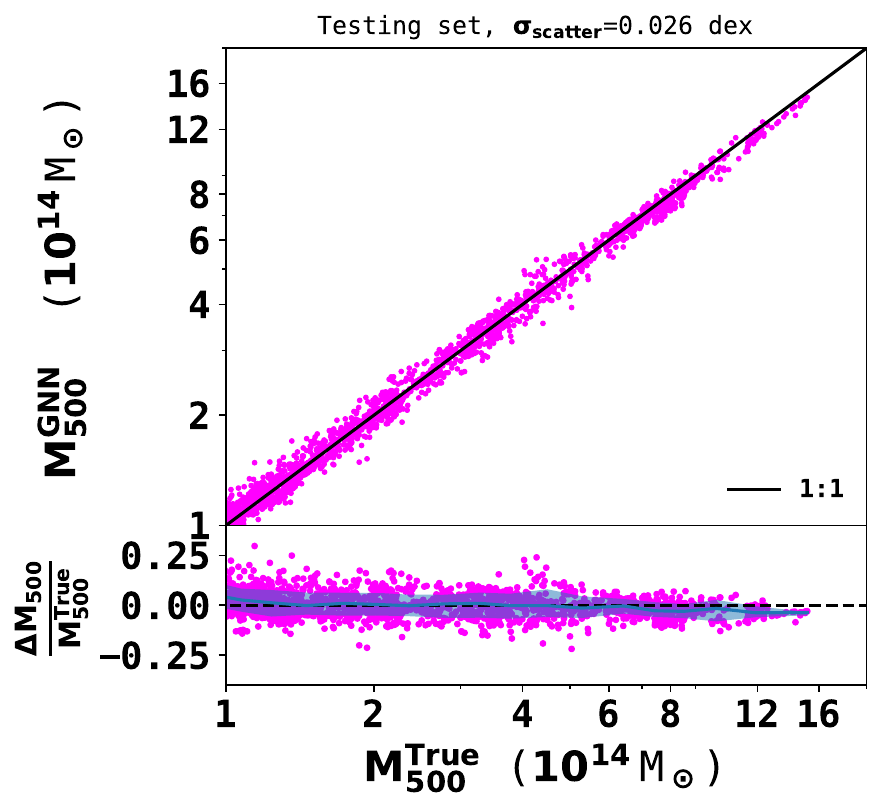}
    \end{minipage}
    \caption{\footnotesize Performance comparison of the lower dimensional 3 feature model (left) and the higher dimensional 6 feature model (right) using the testing sample. The bottom panel in both panels shows the fractional residual distribution with a solid blue line and shaded regions representing the median and 1$\sigma$ dispersion respectively. The lower dimensional model yields a relatively large scatter of 0.038 dex for the testing sample. In contrast, the higher dimensional model shows no improved constraints compared to the fiducial model, with a scatter of 0.026 dex for the testing sample.  }
    \label{fig0appx}
\end{figure*}

Figure~\ref{fig0appx} shows the performances of both models using the testing sample. We find that for the lower dimensional model, the fractional residuals and mass bias to be $0.00\pm0.06$ and $1.00\pm0.07$, respectively, for the testing sample. However, here the scatter is approximately 1.5 times that of our fiducial model. Moreover, for the training sample, we have dispersion comparatively much smaller at 3\% and a scatter of 0.018 dex, indicating overfitting. From the feature importance analysis, we find the relative importance as  [R, $\rho$, T]=[0.54, 0.28, 0.17].

In contrast, the higher dimensional model yields no improved constraints compared to the fiducial case, with a fractional residual of $0.00 \pm 0.06$ and a mass bias of $1.00 \pm 0.05$ for the testing sample. The feature importance distribution for this model is [R, $\rho$, T, P, K, M$_g$] = [0.25, 0.13, 0.03, 0.11, 0.07, 0.39], highlighting that gas mass is still the significant predictor and temperature is the least influential. Even though entropy radially increases, which parallels that of the gas mass, its ability to capture the underlying mass is still poor. Moreover, one must also note that the addition of entropy as a feature does not necessarily improve the model's physical reliability as it is susceptible to observational systematics. In particular, it is known to be strongly affected by gas clumping and small-scale inhomogeneities, which can result in entropy flattening.

\section{Redshift dependence on model training}
\label{app1}
In this Appendix, we investigate the impact of incorporating redshift information during the model training process. Specifically, the model was trained using two distinct redshift snapshots: $z=0.33$ and $z=0.06$, and evaluated on an independent snapshot at $z=0.95$. This setup enabled an assessment of the model’s ability to generalise across different epochs of cosmic time.

A data augmentation strategy, similar to that employed in the development of the fiducial model, was also applied. In this case, however, both the unaugmented training and testing samples were randomly assigned redshifts within the range $0 \leq z \leq 1.2$. By systematically using different redshift bins for the unaugmented samples during training and testing, we explored the model’s sensitivity to cosmic evolution and examined its implications for astrophysical interpretation.

\begin{figure*}
    \centering
    \begin{minipage}{0.45\linewidth}
        \includegraphics[width=\textwidth]{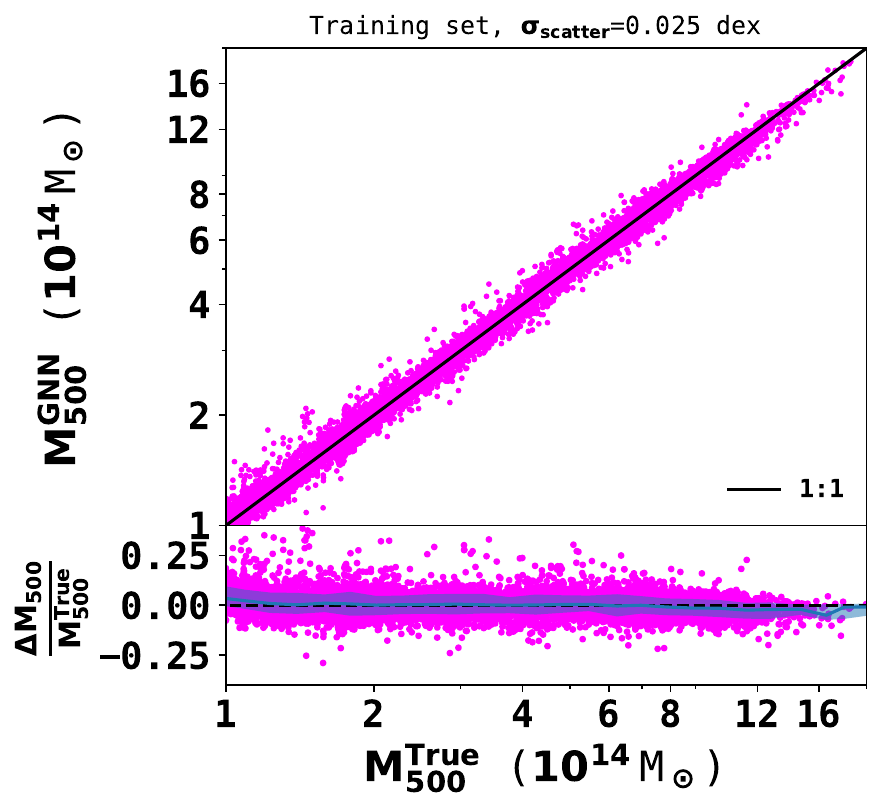} %
    \end{minipage}
    \begin{minipage}{0.45\linewidth}
        \includegraphics[width=\textwidth]{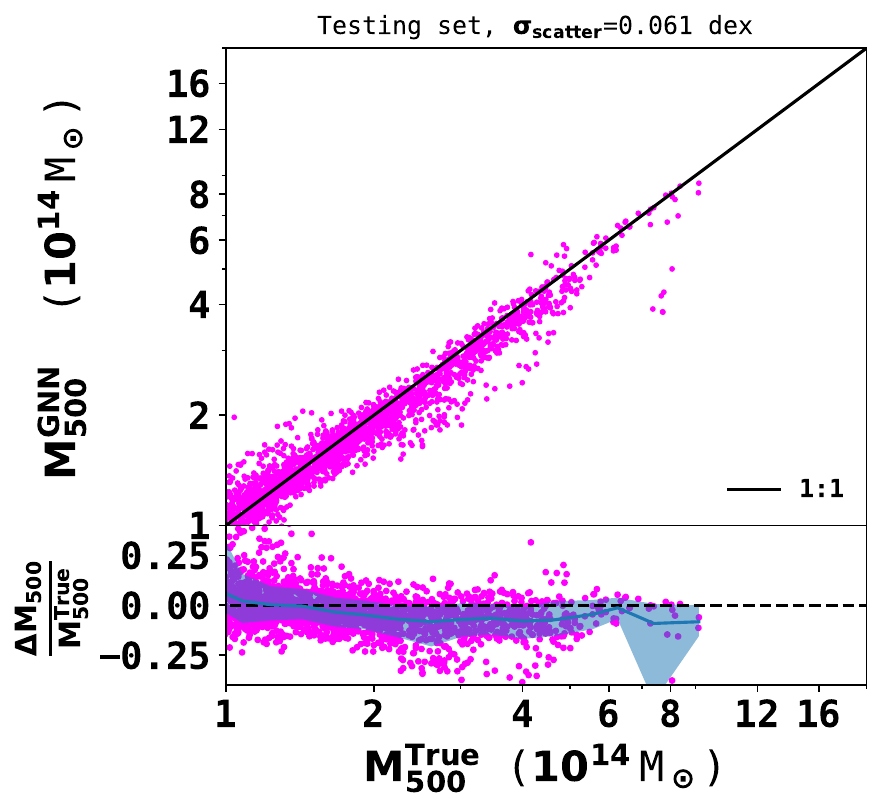}
    \end{minipage}
    \caption{\footnotesize GNN predicted masses, $\text{M}^{\text{GNN}}_{500}$, versus true cluster masses, $\text{M}^{\text{True}}_{500}$, for the training (left) and testing (right) samples, drawn from the original unaugmented dataset across different redshift bins. The black line shows the self-similar relation. The bottom panels show the fractional residuals, with the solid blue line indicating the median and the shaded region representing the 1$\sigma$ scatter. As expected, the model predictions for the training set are effectively unbiased, while the testing sample exhibits a mild average bias of $(1-b)=0.96$, reflecting redshift-dependent effects. }
    \label{fig1appx}
\end{figure*}

Figure~\ref{fig1appx} presents the distribution of GNN predicted cluster masses versus the true cluster masses for both the training and testing samples.  Note that since the testing sample is at high redshift, it contains fewer massive clusters compared to the training sample. As expected, the model’s predictions for the training sample remain effectively unbiased, with predicted values closely following the true cluster masses with a scatter of 0.025 dex. For the testing sample, we observe a bias of  $(1-b) = 0.96 \pm 0.08$ and a scatter of 0.061 dex. While this bias is statistically consistent with unity, it suggests a slight underestimation of cluster masses on average.  This also shows that there is a  significant over-fitting during the training stage. These findings highlight the importance of incorporating redshift-specific information during training to improve model generalisation and ensure accurate mass predictions across a wide span of cosmic time.

\section{Comparison of the M$_{500}^{\textrm{GNN}}$ with M$_{500}^{\textrm{X}}$ and M$_{500}^{\textrm{SZ}}$ masses for REXCESS sample}
\label{app2}
The X-ray derived mass, M\(_{500}^{\textrm{X}} \), is derived from observations linked to the thermal bremsstrahlung emission of the ICM. Here, the observable $Y_{\textrm{X}} =$ M$_g \times$ T is linked to  M\(_{500}^{\textrm{X}} \).
For our work, we consider M$_{500}$ values derived from \cite{2022A&A...665A..24P}. The cluster masses were estimated using the $Y_X$ mass proxy. This proxy is based on the assumption of hydrostatic equilibrium and has been shown to correlate tightly with the total cluster mass. Using XMM-Newton observations of 93 SZE-selected galaxy clusters, they also derived a scaling relation between the core-excised X-ray luminosity ($L_{Xc}$) and the total mass proxy M$_{500}^{\rm X}$, finding a logarithmic intrinsic scatter of approximately 0.13 dex. Table~\ref{appx1} summarises the results of our analysis using the REXCESS sample. 

For the REXCESS sample, we find the following best-fitting relations for the  M\(_{500}^{\textrm{SZ}} \) and M\(_{500}^{\textrm{X}}\) biases with respect to the GNN masses
\begin{eqnarray}
{\textrm M}^{{\text{SZ}}}_{500} = (0.87\pm 0.06) \, {\textrm M}^{\text{GNN}}_{500} +(0.38\pm 0.25) \times 10^{14}{\textrm M}_\odot \nonumber \\ 
{\textrm M}^{{\text{X}}}_{500} = (0.92 \pm 0.02) \, {\textrm M}^{\text{GNN}}_{500} +(0.05\pm 0.0.06) \times 10^{14}{\textrm M}_\odot.
\end{eqnarray}
We also quantify the mass bias \( (1 - b) \) as a function of M\(_{500}^{\text{GNN}} \), yielding
\begin{eqnarray}
(1 - b) = 
\begin{cases}
(0.92 \pm 0.05) \left(\dfrac{\textrm{M}_{500}^{\textrm{GNN}}}{6 \times 10^{14} \, \textrm{M}_\odot}\right)^{-0.17 \pm 0.06}\, \text{M\(_{500}^{\textrm{SZ}}\)}\\ 
(0.93 \pm 0.02) \left(\dfrac{\textrm{M}_{500}^{\textrm{GNN}}}{6 \times 10^{14} \, \textrm{M}_\odot}\right)^{-0.02 \pm 0.23}\, \text{M\(_{500}^{\textrm{X}}\)}

\end{cases}
\end{eqnarray}
We find a mass dependence of bias at around 3$\sigma$ for M$^{\textrm{SZ}}_{500}$-M$^{\textrm{GNN}}_{500}$ relation.
\begin{table*}
\centering
\begin{threeparttable}
\centering
\begin{tabular}{ccccc}
 \toprule
\toprule
Name & redshift & M$_{500}^{{\textrm{X}}}$\tnote{*} & M$_{500}^{{\textrm{SZ}}}$\tnote{**} & M$_{500}^{\textrm{GNN}}$  \\
 & & $10^{14}$M$_\odot$ & $10^{14}$M$_\odot$ & $10^{14}$M$_\odot$\\
\midrule
RXC J0003.8$+$0203& 0.092 & $2.11\pm0.04$ &$2.21\pm0.29$  &$1.96\pm0.11$  \\ \hline 
RXC J0006.0$-$3443& 0.114 & $3.95\pm0.12$ &$3.82\pm0.29$  &$4.77\pm0.25$  \\ \hline 
RXC J0020.7$-$2542& 0.141 & $3.84\pm0.06$ &$4.22\pm0.32$  &$3.70\pm0.21$  \\ \hline 
RXC J0049.4$-$2931& 0.108 & $1.62\pm0.04$ &$2.14\pm0.36$  &$1.60\pm0.10$  \\ \hline 
RXC J0145.0$-$5300& 0.116 & $4.37\pm0.08$ &$3.42\pm0.26$  &$4.64\pm0.23$  \\ \hline 
RXC J0211.4$-$4017& 0.100 & $1.00\pm0.02$ &$1.48\pm0.32$  &$1.03\pm0.09$  \\ \hline 
RXC J0225.1$-$2928& 0.060 & $0.96\pm0.04$ &$0.97\pm0.33$  &$1.16\pm0.10$  \\ \hline 
RXC J0345.7$-$4112& 0.060 & $0.97\pm0.02$ &$1.16\pm0.26$  &$1.10\pm0.09$  \\ \hline
RXC J0547.6$-$3152& 0.148 & $5.01\pm0.10$ &$4.60\pm0.31$  &$5.10\pm0.26$  \\ \hline
RXC J0605.8$-$3518& 0.139 & $3.89\pm0.09$ &$5.11\pm0.27$  &$4.19\pm0.22$  \\ \hline
RXC J0616.8$-$4748& 0.116 & $2.70\pm0.05$ &$2.74\pm0.27$  &$2.82\pm0.16$  \\ \hline
RXC J0645.4$-$5413& 0.164 & $7.38\pm0.18$ &$7.94\pm0.25$  &$8.49\pm0.40$ \\ \hline
RXC J0821.8$+$0112& 0.082 & $1.31\pm0.03$ &$2.27\pm0.30$  &$1.25\pm0.09$  \\ \hline
RXC J0958.3$-$1103& 0.166 & $4.17\pm0.22$ &$5.29\pm0.35$  &$3.98\pm0.20$  \\ \hline
RXC J1044.5$-$0704& 0.134 & $2.69\pm0.02$ &$3.00\pm0.37$  &$2.75\pm0.14$  \\ \hline
RXC J1141.4$-$1216& 0.119 & $2.27\pm0.02$ &$2.80\pm0.36$  &$2.27\pm0.12$  \\ \hline
RXC J1236.7$-$3354& 0.079 & $1.33\pm0.02$ &$   -       $  &$1.29\pm0.10$  \\ \hline
RXC J1302.8$-$0230& 0.084 & $1.89\pm0.03$ &$2.15\pm0.39$  &$1.81\pm0.10$  \\ \hline
RXC J1311.5$-$0120& 0.183 & $8.41\pm0.08$ &$8.72\pm0.35$  &$8.42\pm0.39$  \\ \hline
RXC J1516.3$+$0005& 0.118 & $3.28\pm0.07$ &$2.19\pm0.58$  &$3.38\pm0.18$  \\ \hline
RXC J1044.5$-$0704& 0.119 & $2.59\pm0.05$ &$3.76\pm0.41$  &$3.00\pm0.16$  \\ \hline
RXC J2014.8$-$2430& 0.153 & $5.38\pm0.07$ &$5.14\pm0.35$  &$5.95\pm0.31$  \\ \hline
RXC J2023.0$-$2056& 0.056 & $1.21\pm0.03$ &$0.84\pm0.40$  &$1.31\pm0.09$  \\ \hline
RXC J2048.1$-$1750& 0.147 & $4.31\pm0.07$ &$4.25\pm0.36$  &$5.20\pm0.27$  \\ \hline
RXC J2129.8$-$5048& 0.079 & $2.26\pm0.06$ &$1.91\pm0.27$  &$2.61\pm0.14$  \\ \hline
RXC J2149.1$-$3041& 0.118 & $2.25\pm0.03$ &$1.94\pm0.39$  &$2.27\pm0.12$  \\ \hline
RXC J2157.4$-$0747& 0.057 & $1.29\pm0.03$ &$1.46\pm0.29$  &$1.53\pm0.10$  \\ \hline
RXC J2217.7$-$3543& 0.148 & $3.61\pm0.05$ &$4.52\pm0.33$  &$3.96\pm0.20$  \\ \hline
RXC J2218.6$-$3853& 0.141 & $4.92\pm0.11$ &$4.01\pm0.34$  &$5.16\pm0.27$  \\ \hline
RXC J2234.5$-$3744& 0.151 & $7.36\pm0.09$ &$6.87\pm0.27$  &$8.11\pm0.37$  \\ \hline
RXC J2319.6$-$7313& 0.098 & $1.56\pm0.03$ &$1.78\pm0.29$  &$1.59\pm0.10$  \\
\bottomrule
\end{tabular}
\begin{tablenotes}
\item[*] Derived from \cite{2022A&A...665A..24P}.
\item[**] Derived as detailed in Sect.~\ref{sec:szmass}.
\end{tablenotes}
\caption{\footnotesize Results with REXCESS sample. Columns: (1) Cluster name, (2) Redshift, (3) Mass obtained using from $Y_{\textrm{ X}}$-M$_{500}^{\textrm{X}}$,(4) GNN estimated mass, M$_{500}^{\textrm {GNN}}$  (this work).} \label  {appx1}
\end{threeparttable}
    \end{table*}

\section{Comparison of the M$_{500}^{\textrm{GNN}}$ with hydrostatic and M$_{500}^{\textrm{SZ}}$ masses for X-COP sample}
\label{app3}
\cite{2019A&A...621A..40E} not only estimated the hydrostatic masses (M$_{500}^{\textrm{HSE}}$) but also proposed a method to estimate the total mass which includes the contribution from non-thermal pressure \citep{2014ApJ...792...25N} (M$_{500}^{\textrm{HSE+NT}}$). They considered a functional form for the non-thermal pressure derived from hydrodynamical simulations along with the hydrostatic approximation while assuming a universal gas mass fraction at the varial radius . Table~\ref{appx2} summarises the comparison of cluster mass estimates for the X-COP sample derived using various methodologies.

For the X-COP sample, we find the following best-fitting relations for the  M\(_{500}^{\textrm{SZ}} \), M\(_{500}^{\textrm{HSE}} \) and M\(_{500}^{\textrm{HSE+NT}} \) biases with respect to the GNN masses
\begin{eqnarray}
{\textrm M}^{{\text{SZ}}}_{500} = (0.61 \pm 0.06) \, {\textrm M}^{\text{GNN}}_{500} +(0.93\pm 0.50) \times 10^{14}{\textrm M}_\odot \nonumber \\ 
{\textrm M}^{{\text{HST}}}_{500} = ( 0.51 \pm 0.17) \, {\textrm M}^{\text{GNN}}_{500} +(2.32\pm 1.30) \times 10^{14}{\textrm M}_\odot \nonumber \\ 
{\textrm M}^{{\text{HST+NT}}}_{500} = (0.88 \pm 0.12) \, {\textrm M}^{\text{GNN}}_{500} +(0.26\pm 1.02) \times 10^{14}{\textrm M}_\odot.
\end{eqnarray}
The mass bias \( (1 - b) \) as a function of M\( _{500}^{\text{GNN}} \) for the X-COP sample gives
\begin{eqnarray}
(1 - b) = 
\begin{cases}
(0.87 \pm 0.03) \left(\dfrac{\textrm{M}_{500}^{\textrm{GNN}}}{6 \times 10^{14} \, \textrm{M}_\odot}\right)^{-0.18 \pm 0.09}, & \text{M\(_{500}^{\textrm{SZ}}\)} \\
(0.87 \pm 0.07) \left(\dfrac{\textrm{M}_{500}^{\textrm{GNN}}}{6 \times 10^{14} \, \textrm{M}_\odot}\right)^{-0.28 \pm 0.20}, & \text{M\(_{500}^{\textrm{HST}}\)} \\
(0.92 \pm 0.05) \left(\dfrac{\textrm{M}_{500}^{\textrm{GNN}}}{6 \times 10^{14} \, \textrm{M}_\odot}\right)^{-0.02 \pm 0.14}, & \text{M\(_{500}^{\textrm{HST+NT}}\)}
\end{cases}
\end{eqnarray}
For the X-COP sample, we find weak evidence, at the 2$\sigma$ level,  for  a mass dependent bias in the SZ masses relative to the GNN masses.

\begin{table*}
\centering
\begin{threeparttable}

\begin{tabular}{c c c c c c}
 \toprule
\toprule
Name & redshift & M$_{500}^{\textrm{SZ}}$\tnote{*} & M$_{500}^{\textrm{HSE}}$\tnote{**}& M$_{500}^{\textrm{HSE+NT}}\tnote{**}$& M$_{500}^{\textrm{GNN}}$  \\
& &  $10^{14}$M$_\odot$ & $10^{14}$M$_\odot$ &  $10^{14}$M$_\odot$ &  $10^{14}$M$_\odot$   \\
\midrule
A85     &   0.055   &     $ 4.80 \pm 0.16$ &     $5.65 \pm0.18$   & $ 6.22 \pm  0.49$    &   $  6.79 \pm 0.33$  \\  \hline
A644    &   0.070   &     $ 5.10 \pm 0.18$ &     $5.66 \pm0.48$   & $6.03 \pm    0.65    $ & $  6.25 \pm 0.31$  \\  \hline 
A1644   &   0.047   &     $ 3.43 \pm 0.17$ &     $3.48 \pm0.20$   & $3.52 \pm    0.21    $ & $  3.82 \pm 0.20$  \\ \hline  
A1795   &   0.062   &     $ 4.47 \pm 0.16$ &     $4.53 \pm0.14$   & $4.77 \pm    0.33    $ & $  5.25 \pm 0.27$  \\ \hline   
A2029   &  0.076    &     $ 7.20 \pm 0.20$ &     $8.65 \pm0.29$   & $8.98 \pm    0.83    $ & $  9.63 \pm 0.46$  \\ \hline   
A2142   &   0.090   &     $ 8.88 \pm 0.21$ &     $8.95 \pm0.26$   & $10.05\pm    0.73    $ & $  11.50\pm 0.46$  \\ \hline   
A2255   &   0.080   &     $ 5.20 \pm 0.14$ &     $5.26 \pm0.34$   & $ 5.87 \pm   0.46    $ & $  7.48 \pm 0.35$  \\ \hline   
A2319   &   0.055   &     $ 8.72 \pm 0.13$ &     $7.31 \pm0.28$   & $11.44\pm    1.08    $ & $  13.97\pm 0.39$ \\ \hline    
A3158   &   0.059   &     $ 4.12 \pm 0.15$ &     $4.26 \pm0.18$   & $4.53 \pm    0.37    $ & $  5.10 \pm 0.27$ \\ \hline    
A3266   &   0.058   &     $ 6.56 \pm 0.12$ &     $8.80 \pm0.57$   & $8.94 \pm    0.56    $ & $  8.89 \pm 0.42$ \\ \hline    
RXC1825 &   0.065   &     $ 3.71 \pm 0.19$ &     $4.08 \pm0.13$   & $3.94 \pm    0.32    $ & $  5.20 \pm 0.27$ \\ \hline   
ZW1215  &   0.076   &     $ 4.04 \pm 0.21$ &     $7.66 \pm0.52$   & $7.67 \pm    0.53    $ & $  6.02 \pm 0.32$ \\

\midrule
\end{tabular}
\begin{tablenotes}
\item[*] Derived as detailed in Sect.~\ref{sec:szmass}.
\item[**] Derived from \cite{2019A&A...621A..40E}.
\end{tablenotes}
\caption{\footnotesize Results with X-COP sample. Columns: (1) Cluster name, (2) Redshift, (3) Mass obtained using from $Y_{\textrm{ SZ}}$-M$_{500}^{\textrm{SZ}}$, (4) Hydrostatic mass, M$_{500}^{\textrm {HSE}}$, (5) Hydrostatic mass corrected with non-thermal pressure, M$_{500}^{\textrm{ HSE+NT}}$, (6) GCN estimated mass, M$_{500}^{\textrm {GCN}}$ (this work).} \label  {appx2}
\end{threeparttable}
    \end{table*}

\end{document}